 \documentclass[final,3p,times,twocolumn]{elsarticle}

\usepackage{amssymb}

\usepackage{lineno}

\usepackage{hyperref}
\usepackage{color}
\newcommand{\Aprime}{A\ensuremath{^\prime}}
\newcommand{\ee}{e$^+$e$^-$}
\newcommand{\fluenceunit}{1~MeV~neutron~equivalent/cm\ensuremath{^2}}

\newcommand{\egs}{{\sc EGS5}}

\journal{Nuclear Instruments and Methods in Physics Research Section A}

\begin{document}

\begin{frontmatter}



\title{The Heavy Photon Search test detector}


\newcommand{\red[1]}{{\color{red}{\bf #1}}}
\newcommand{\JLAB}{Thomas Jefferson National Accelerator Facility, Newport News, Virginia 23606}
\newcommand{\CUA}{Catholic University of America, Washington, D.C. 20064}
\newcommand{\OU}{Ohio University,  Athens, Ohio 45701}
\newcommand{\YEREVAN}{Yerevan Physics Institute, 375036 Yerevan, Armenia}
\newcommand{\SCAROLINA}{University of South Carolina, Columbia, South Carolina 29208}
\newcommand{\NSU}{Norfolk State University, Norfolk, Virginia 23504}
\newcommand{\ODU}{Old Dominion University, Norfolk, Virginia 23529}
\newcommand{\genova}{Istituto Nazionale di Fisica Nucleare, Sezione di Genova e Dipartimento di Fisica dell\'Universita, 16146 Genova, Italy}
\newcommand{\SACLAY}{CEA, Centre de Saclay, Irfu/Service de Physique Nucl\'eaire 91191 Gif-sur-Yvette, France}
\newcommand{\ORSAY}{Institut de Physique Nucl\'eaire, CNRS/IN2P3 and Universit\'e Paris Sud, Orsay, France}
\newcommand{\ECOLE}{CPhT, Ecole Polytechnique, F 91128 PALAISEAU CEDEX, France}
\newcommand{\UCSC}{Santa Cruz Institute for Particle Physics, University of California, Santa Cruz, CA 95064}
\newcommand{\SUNY}{Stony Brook University, Stony Brook, NY 11794-3800}
\newcommand{\FNAL}{Fermi National Accelerator Laboratory, Batavia, IL 60510-5011}
\newcommand{\UNH}{University of New Hampshire, Department of Physics, Durham, NH 03824}
\newcommand{\PERIMETER}{Perimeter Institute, Ontario, Canada N2L 2Y5}
\newcommand{\RPI}{Rensselaer Polytechnic Institute, Department of Physics, Troy, NY 12181}
\newcommand{\SLAC}{SLAC National Accelerator Laboratory, Menlo Park, CA 94025}
\newcommand{\WNM}{The College of William and Mary, Department of Physics, Williamsburg, VA 23185}
\newcommand{\GLASGOW}{University of Glasgow, Glasgow, G12 8QQ, Scotland, UK}
\newcommand{\RUTGERS}{Rutgers University, Department of Physics and Astronomy, Piscataway, NJ 08854}

\author[GENOVA]{M. Battaglieri} 
\author[JLAB]{S. Boyarinov} 
\author[ODU]{S. Bueltmann} 
\author[JLAB]{ V. Burkert} 
\author[GENOVA]{A. Celentano} 
\author[ORSAY]{ G. Charles}
\author[FNAL]{W. Cooper}
\author[JLAB]{C. Cuevas}
\author[YEREVAN]{N. Dashyan} 
\author[GENOVA]{ R. DeVita}
\author[ORSAY]{ C. Desnault}
\author[JLAB]{ A. Deur} 
\author[JLAB]{ H. Egiyan}
 \author[JLAB]{ L. Elouadrhiri} 
\author[SUNY]{R. Essig}
\author[UCSC]{ V. Fadeyev} 
\author[SLAC]{C. Field}
 \author[JLAB]{ A. Freyberger} 
\author[RUTGERS]{Y. Gershtein}
\author[YEREVAN]{ N. Gevorgyan} 
 \author[JLAB]{ F.-X. Girod} 
\author[SLAC]{N. Graf} 
\author[SLAC]{ M. Graham} 
\author[WNM]{K. Griffioen}
\author[UCSC]{A. Grillo} 
\author[ORSAY]{ M. Guidal} 
\author[SLAC]{ G. Haller} 
\author[SLAC]{P. Hansson Adrian\corref{corrauthor}}
\ead{phansson@slac.stanford.edu}
\author[SLAC]{ R. Herbst} 
\author[UNH]{M. Holtrop}
\author[SLAC]{ J. Jaros}
 \author[JLAB]{ S. Kaneta}
\author[NSU]{M. Khandaker} 
\author[RPI]{A. Kubarovsky}
\author[JLAB]{ V. Kubarovsky} 
\author[SLAC]{T. Maruyama} 
\author[SLAC]{ J. McCormick} 
\author[SLAC]{ K. Moffeit} 
\author[UCSC]{ O. Moreno}
\author[SLAC]{ H. Neal} 
\author[SLAC]{ T. Nelson} 
\author[ORSAY]{ S. Niccolai} 
\author[SLAC]{ A. Odian} 
\author[SLAC]{ M. Oriunno} 
\author[YEREVAN]{ R. Paremuzyan}
\author[SLAC]{ R. Partridge} 
\author[UNH]{ S. K. Phillips}
\author[ORSAY]{ E. Rauly}
 \author[JLAB]{ B. Raydo} 
\author[RUTGERS]{J. Reichert}
\author[ORSAY]{ E. Rindel} 
\author[ORSAY]{ P. Rosier}
\author[NSU]{ C. Salgado}
\author[PERIMETER]{P. Schuster} 
\author[JLAB]{ Y. Sharabian} 
\author[GLASGOW]{D. Sokhan}
\author[JLAB]{ S. Stepanyan}
\author[PERIMETER]{ N. Toro}
\author[SLAC]{ S. Uemura} 
\author[JLAB]{ M. Ungaro} 
\author[YEREVAN]{ H. Voskanyan}
\author[SLAC]{ D. Walz}
\author[ODU]{ L. B. Weinstein}
\author[JLAB]{ B. Wojtsekhowski}

\address[GENOVA]{\genova}
\address[JLAB]{\JLAB}
\address[ODU]{\ODU}
\address[FNAL]{\FNAL}
\address[YEREVAN]{\YEREVAN}
\address[ORSAY]{\ORSAY}
\address[SUNY]{\SUNY}
\address[UCSC]{\UCSC}
\address[SLAC]{\SLAC}                                 
\address[RUTGERS]{\RUTGERS}
\address[WNM]{\WNM}
\address[UNH]{\UNH}
\address[NSU]{\NSU}
\address[RPI]{\RPI}
\address[PERIMETER]{\PERIMETER}
\address[GLASGOW]{\GLASGOW}

\cortext[corrauthor]{Corresponding author.}



\begin{abstract}
The  Heavy Photon Search (HPS), an experiment to search for a hidden sector photon in fixed target 
electroproduction, is preparing for installation at the Thomas Jefferson National Accelerator Facility 
(JLab) in the Fall of 2014.  As the first stage of this project, the HPS Test Run apparatus was 
constructed and operated in 2012 to demonstrate the experiment's technical feasibility and to  confirm 
that the trigger rates and occupancies are as expected. This paper describes the HPS Test Run 
apparatus and readout electronics and its performance. In this setting, a heavy photon can be identified 
as a narrow peak in the \ee{} invariant mass spectrum above the trident background or as a narrow  
invariant mass peak with a decay vertex displaced from the production target, so charged particle 
tracking and vertexing are needed for its detection. 
In the HPS Test Run, charged particles are measured with a compact forward silicon microstrip tracker 
inside a dipole magnet. Electromagnetic showers are detected in a PbW0$_{4}$ 
crystal calorimeter situated behind the magnet,  and are used to trigger the experiment and identify 
electrons and positrons. Both detectors are placed close to the 
beam line and split top-bottom. This arrangement provides sensitivity to low-mass heavy photons,  
allow clear  passage of the unscattered beam, 
and avoids the spray of degraded electrons coming from the target. 
The discrimination between prompt and displaced \ee{} pairs requires the first layer of silicon 
sensors be placed only 10~cm downstream of the target. The expected signal is small, and the 
trident background huge, so the experiment requires very large statistics. Accordingly, the
HPS Test Run utilizes  high-rate readout and data acquisition electronics and a fast trigger to exploit 
the essentially 100\% duty cycle of the CEBAF accelerator at JLab.
\end{abstract}


\end{frontmatter}

 
\newpage


\section{Introduction}
\label{introduction}
The heavy photon (\Aprime{}), aka a ``hidden sector'' or ``dark'' photon, is a massive gauge boson which 
couples weakly to electric charge by mixing with the Standard Model 
photon~\cite{Holdom:1985ag,Galison:1983pa}. Consequently, it can be radiated by electrons and subsequently decay 
into \ee{} pairs, albeit at rates far below those of QED trident processes. Heavy photons have been 
suggested by numerous beyond Standard Model theories~\cite{Essig:2013lka} to explain the 
discrepancy 
between theory and experiment of the muon's $g-2$ ~\cite{Pospelov:2008zw}, and as a possible 
explanation of recent astrophysical 
anomalies, e.g.~\cite{Adriani:2008zr,FermiLAT:2011ab,Aguilar:2013qda}.  
Heavy photons couple directly to  hidden sector particles with ``dark'' or ``hidden sector'' charge; these 
particles could constitute all or some of the dark matter, e.g.~\cite{ArkaniHamed:2008qn,Pospelov:2008jd}. 
Current phenomenology highlights the $20-1000$~MeV/c$^{2}$ mass range, and 
suggests that  the coupling to electric charge, $\epsilon e$, has $\epsilon$ in the range of 
$10^{-3} -10^{-5}$. This range of parameters makes \Aprime{} searches viable in 
medium energy fixed target electroproduction~\cite{Bjorken:2009mm}, but requires large data sets and 
good mass resolution to identify a small mass peak above the copious QED background. At small 
couplings, the \Aprime{} becomes long-lived, so detection of a displaced decay vertex can reject the prompt 
QED background and boost experimental sensitivity.  

The HPS experiment~\cite{HPS_proposal_2010} is preparing for installation in Hall-B at JLab in the Fall of 
2014 to search for heavy photons by directing the 2.2-6.6~GeV CEBAF12 electron beam onto a thin 
(0.25\% $X_{0}$) Tungsten target foil. The HPS experiment uses both 
invariant mass and secondary vertex signatures to search for \Aprime{} decays into \ee{} pairs. 
At CEBAF energies, the \Aprime{} decay products are boosted along the beam axis with small opening angles.
For couplings $\epsilon << 10^{-3}$,  \Aprime{} decay lengths range from millimeters to tens of centimeters and 
beyond. Accordingly the tracking detectors cover opening angles down to 15~mrad and are placed just 10~cm 
downstream of the target.

HPS employs a 90~cm long silicon tracking and vertexing detector located inside a dipole magnet to measure 
momenta and decay vertex positions. A fast PbWO$_{4}$ electromagnetic calorimeter downstream of the 
magnet provides the trigger and electron identification. Both the silicon tracker and the ECal have $\sim$ns 
timing resolution, which eliminates much of the out-of-time background from multiple scattered beam electrons.  
Fast front end electronics and high trigger and data rate capability and the effectively 100\% duty cycle of the 
CEBAF accelerator allows HPS to accumulate the very large statistics needed to be sensitive to the highly 
suppressed production of heavy photons.

The HPS Test Run, using a simplified version of the HPS apparatus, was proposed and 
approved at JLab as the first stage of HPS. 
Its purposes included demonstrating that the apparatus and data acquisition systems are 
technically feasible and  the trigger rates and occupancies to be encountered in electron-beam 
running are as simulated. Given dedicated running time with electron beams, the HPS Test Run 
apparatus is capable of searching for heavy photons in unexplored regions of parameter space.    
Therefore, key design criteria and requirements for HPS and the HPS Test Run apparatus are the same:
\begin{itemize}
\item uniform acceptance between 15 and approximately 70~mrad in the forward region to catch boosted decay products close to the beam, 
\item beam passage through the apparatus in vacuum, to eliminate direct interactions with the 
detector and minimize beam gas interactions, 
\item detector components that can survive and efficiently operate in a high radiation environment 
with some localized doses at the 100~Mrad level,
\item high-rate electronics, handling trigger rates up to 50~kHz and data rates of 100~MB/s to permanent 
storage, 
\item a flexible, redundant and efficient trigger for selecting electron and positron pairs,
capable of handling rates up to 50~kHz,
\item hit reconstruction efficiency higher than 99\% and average track reconstruction efficiency higher than 98\% for electrons and positrons,
\item 2~ns hit time resolution in the silicon vertex tracker, 
\item A' mass resolution of 2.5\% or better, which translates to momentum resolution of 4.5\% and angular 
resolution of 2~mrad/$p$(GeV/c)) for $B$$=$0.5~T,
\item resolution of distance of closest approach to the beam axis less than 250 (100)~$\mu$m for tracks with 
0.5 (1.7)~GeV/c. This gives a decay length resolution of about 1~mm for a 100~MeV/c$^2$ \Aprime{}.
\item PbWO$_{4}$ electromagnetic calorimeter energy resolution $\Delta E/E$$\leq$$5\%$/$\sqrt{E}$ and transverse
segmentation $\sim$1.5~cm ($<$ Moliere radius in PbWO$_{4}$).  The energy resolution requirement for
triggering is less stringent because the electrons and positrons have $E > 0.5$~GeV. The segmentation provides good spatial resolution and guarantees minimal shower overlap with background hits.
\end{itemize}

The HPS Test Run apparatus was installed on April 19, 2012, and ran parasitically in the photon beam of the HDice 
experiment~\cite{HDice} until May 18. The JLab run schedule 
precluded any dedicated electron beam running, but the HPS Test Run was allowed an eight hour 
dedicated photon beam run at the end of scheduled CEBAF running. During this dedicated 
period, \ee{} pairs, produced in a gold foil upstream of the experiment, were studied. With no dedicated 
electron beam running, it was not possible to search for an A'. However, the final running
provided enough data to demonstrate the functionality of the apparatus, document its performance, and 
explore trigger rates, as shown below.

This paper reviews the HPS Test Run apparatus, documenting the performance of the trigger, 
data acquisition, silicon tracking and vertex detector, and the electromagnetic calorimeter at, or close 
to, the level required for the HPS experiment.

\section{Detector overview}
\label{detector}
{\small
\begin{figure*}[t]
\begin{center}
    \includegraphics[width=0.8\textwidth]{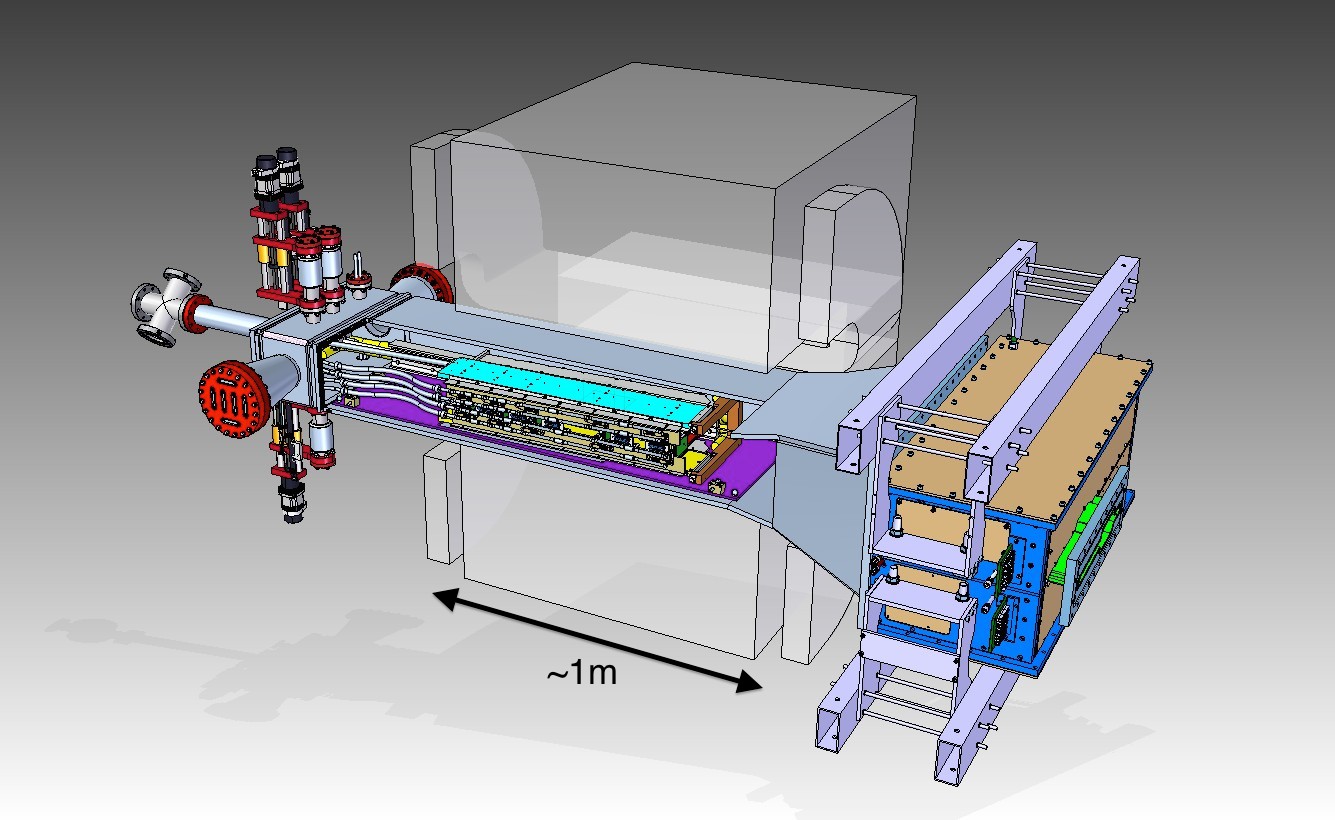}
\caption{Rendering of the HPS Test Run apparatus installed on the beam line.
\label{fig:testrundetector}}
\end{center}
\end{figure*}
}
The HPS Test Run apparatus was designed to run in Hall~B at JLab using the CEBAF 499~MHz 
electron beam at energies between 2.2 and 6.6~GeV and currents between 200 and 600~nA.  
The overall design of the experiment follows from the kinematics of \Aprime{} production which 
typically results in a final state particle within a few degrees of the incoming beam, especially at low 
$m_{\Aprime{}}$. Detectors must therefore be placed close to the beam. 
The intense electron beam enlarges downstream after multiple scattering in the target and electrons 
which have radiated in the target disperse horizontally in the field of the analyzing magnet. Together 
they constitute a ``wall of flame'' which must be completely avoided. Accordingly, 
the apparatus is split vertically to avoid a ``dead zone'', the region within $\pm 15$~mrad of the beam 
plane. In addition, the beam is transported in vacuum through the tracker to minimize beam-gas 
interaction backgrounds. Even with these precautions, the occupancies of sensors near the beam 
plane are high, dominated by the
multiple Coulomb scattering of the primary beam, so high-rate detectors, a fast trigger, 
and excellent time tagging are required to minimize their impact.  
The trigger comes from a highly-segmented  lead-tungstate (PbWO$_{4}$) crystal calorimeter located just 
downstream of the dipole magnet. 
\begin{center}
\begin{table*}[t]
{\small
\caption{Overview of the coverage, segmentation and performance of the HPS Test Run detector. The $\sigma_{d_0}$ is the track impact parameter resolution of the SVT at the nominal oron target position.  $\sigma_{pos}$ is the estimated position resolution perpendicular to the strip direction on the silicon sensors of the SVT.
\label{tab:detector-overview}}
\begin{tabular}{lccccccc}
\hline 
System & Coverage & \# channels & ADC & \# layers & Segmentation & Time resolution  & Performance \\
 & (mrad) &  & (bit) & &  & (ns)  &  \\
\hline
SVT & $15<\theta_{y} < 70$ & 12780 & 14 & 5    & $30~\mu$m (sense) & $2.5$& $\sigma_{d0,y}  \approx 100~\mu$m \\
& (5 hits) &  &  & (stereo layers) & $60~\mu$m (readout)  & & $\sigma_{d0,x} \approx 300~\mu$m \\
& &  &  &  & ($\sigma_{pos} \approx 6~\mu$m) &  & $\sigma_{d0,z}\approx 1$~mm \\
\hline
ECal & $15<\theta_{y} < 60$ & 442 & 12  & 1  & $1.33\times1.33$~cm$^2$ & 4  & $\sigma(E)/E \approx 4.5\%/sqrt{E}$ \\ 
 &  &  &   &  & $1.6\times1.6$~cm$^2$  & (trigger)& Ref.~\cite{clas_ecal,clas_ecal2,clas_thesis}  \\ 
\hline
\end{tabular}
}
\end{table*}
\end{center}
A rendering of the apparatus installed on the beam line is shown in 
Figure~\ref{fig:testrundetector} and an overview of the coverage, segmentation and performance is 
given in Table~\ref{tab:detector-overview}.  

The silicon tracking and vertexing detector for the HPS Test Run, or SVT, resides in a vacuum 
chamber inside the Pair Spectrometer (PS) dipole magnet in Hall B at JLab. The magnetic field 
strength was 0.5~T oriented vertically throughout the run. The SVT has five 
measurement stations, or ``layers,'' beginning 10 cm downstream of the target. Each layer 
comprises a pair of closely-spaced silicon microstrip sensors responsible for measuring a single 
coordinate, or ``view''. Introduction of a small (50 or 100~mrad) stereo angle between the two 
sensors of each layer provides three-dimensional tracking and vertexing throughout the acceptance 
of the detector. In order to accommodate the dead zone, the 
SVT is built in two halves that are approximately mirror reflections of one another about the plane of the 
nominal electron beam.  Each layer in one half is supported on a common support plate with 
independent cooling and readout. 

The electromagnetic calorimeter (ECal) is also split into two halves. Each half of the ECal consists of 
221 PbWO$_4$ crystals arranged in rectangular formation. There are five rows with 46 crystals in 
each row except the row closest to the beam plane which has 37. The light from each crystal 
is read out by an Avalanche Photodiode (APD) glued on the back surface of the crystal. 
Signals from the APDs are amplified using custom-made amplifier boards before being sent to the 
data acquisition electronics.

The  Data Acquisition system combines two architectures, the Advanced Telecom Communications 
Architecture (ATCA) based SVT readout system and VMEbus Switched Serial (VXS) based digitization 
and triggering system for the ECal.


\section{The HPS Test Run beamline}
\begin{figure}[]
    \includegraphics[width=0.45\textwidth]{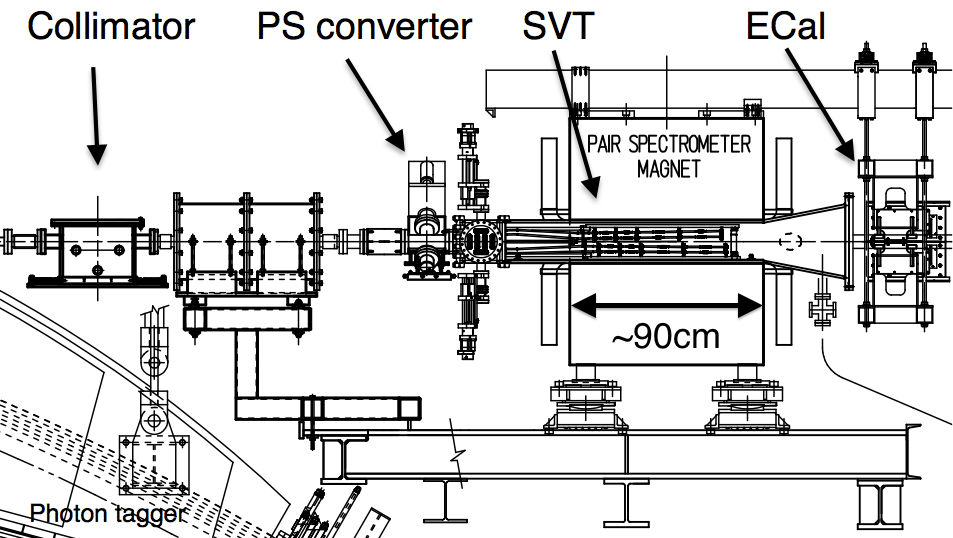}
\caption{\small{Layout of the HPS parasitic run.}
\label{fig:hpstest_layout}}
\end{figure}

Since an electron beam was unavailable, the HPS Test Run detected the electrons and positrons 
produced by interactions of the secondary photon beam with a thin foil just upstream of the detectors.  
The HPS Test Run studied the performance of the detectors and the multiple Coulomb scattering of the 
electrons and positrons. Figure~\ref{fig:hpstest_layout} shows the layout of 
the setup on the beam line. The SVT was installed inside the Hall B pair 
spectrometer magnet vacuum chamber with the ECal mounted downstream of it. Both the 
SVT and the ECal were retracted off the beam plane compared to nominal electron beam running to 
allow clean passage of the photon beam through the system. 

The photon beam was generated in the interaction of $5.5$~GeV electrons with a $10^{-4}~X_0$ 
gold radiator located $\approx~9$~m upstream of the PS. The primary beam and scattered 
electrons are deflected away from detectors by the dipole magnet of the photon tagging system. 
During the dedicated HPS Test Run period, the collimated (6.4~mm diameter) photon beam passes 
through the PS pair converter gold foil and later the HPS system.
The PS pair converter was located $\approx 77$~cm upstream of the first layer of the SVT.

 Data was taken on three different converter thicknesses with photon fluxes between $0.4$-$1.3\times10^8$/s at photon energies between 0.55 and 5.5 GeV produced by a 30-90 nA electron 
 beam. Data was measured for both polarities of the PS dipole magnet.
\begin{table}[]
\begin{center}
{\small
\begin{tabular}{|c|c|c|}
\hline
Converter thickn. & Duration &  $e^-$ on radiator \\
 (\%$X_0$) & (s) & ($\mu$C)    \\   
\hline
0    & 1279  &   88.1 \\ 
0.18   & 2640 &   193.5 \\ 
0.45  & 2149 &     140.7 \\ 
1.6   & 911 &   24.4 \\ 
\hline
\end{tabular}
}
\caption{Measured integrated currents for the dedicated photon runs.
\label{tab:currents}}
\end{center}
\end{table}
The photon beam line during the HPS Test Run produced a relatively large number of \ee{} pairs 
originating upstream of the converter position. This contribution was measured during data taking 
with ``empty'' converter runs, i.e. removing the converter but with all other conditions the same. 
The runs taken during the time dedicated to HPS Test Run are summarized in Table~\ref{tab:currents}.


\section{Silicon Vertex Tracker}
\label{svt}

The Silicon Vertex Tracker (SVT) enables efficient reconstruction of charged particles and precise 
determination of their trajectories. This allow \Aprime{} decays to be distinguished 
from background via simultaneous measurements of the invariant mass of \ee{} decay products and the 
position of decay vertices downstream of the target. 

The design of the SVT is primarily driven by physics requirements and constraints from the 
environment at the interaction region. 
The \Aprime{} decay products have momenta in the range of 0.4-2.0~GeV$/$c (for a 2.2~GeV beam), so 
multiple scattering 
dominates mass and vertexing uncertainties for any possible material budget. The SVT must therefore 
minimize the amount of material in the tracking volume. 
The signal yield for long-lived \Aprime{} is very small, so 
the rejection of prompt vertices must be exceedingly pure, on the order of $10^{-7}$, in order to 
eliminate all prompt backgrounds. To achieve the required vertexing performance the first layer of the 
SVT must be placed no more than about 10~cm downstream of the target. At that distance, it is found 
that the active region of a sensor can be placed as close as 1.5~mm from the center of the beam, 
defining the 15~mrad ``dead zone'' mentioned previously, to maximize low-mass \Aprime{} acceptance 
with decay products nearly collinear with the beam axis. At the edge of this ``dead zone'', the 
radiation dose approaches $10^{15}$~electrons/cm$^2$/month, or roughly 
$3 \times 10^{13}$~\fluenceunit{}/month~\cite{Rashevskaya:2002nd}, 
requiring the sensors to be actively cooled. 
Meanwhile, very low-energy delta rays from 
beam-gas interactions would multiply the density of background hits, so the SVT must operate inside the 
beam vacuum.  Finally, in order to protect the sensors, the detector must be movable so that it can be 
retracted during periods of uncertain beam conditions or beam tuning.  

A mass resolution of 2.5\% is adequate to extend a bump-hunt search for an \Aprime{} into virgin territory. 
For running at 2.2~GeV, this translates into a requirement for track momentum ($p$) resolution of 4-5\% and 
angular resolution of about 2~mrad/$p$(GeV/c)~\cite{HPS_proposal_2010}. 
Multiple Coulomb scattering dominates both the mass and vertexing uncertainties, relaxing 
the spatial hit resolution requirement to $<100~\mu$m (50~$\mu$m) in the bend (non-bend) plane.

High background occupancies, up to 4~MHz$/$mm$^{2}$ locally,  in the region closest to the beam result 
from beam electrons undergoing multiple scattering in the target. These background hits are rejected by  
requiring reconstruction of the hit time relative to the trigger with 2~ns resolution.

\subsection{Layout}
\begin{center}
\begin{figure}[htp]
\includegraphics[width=7cm]{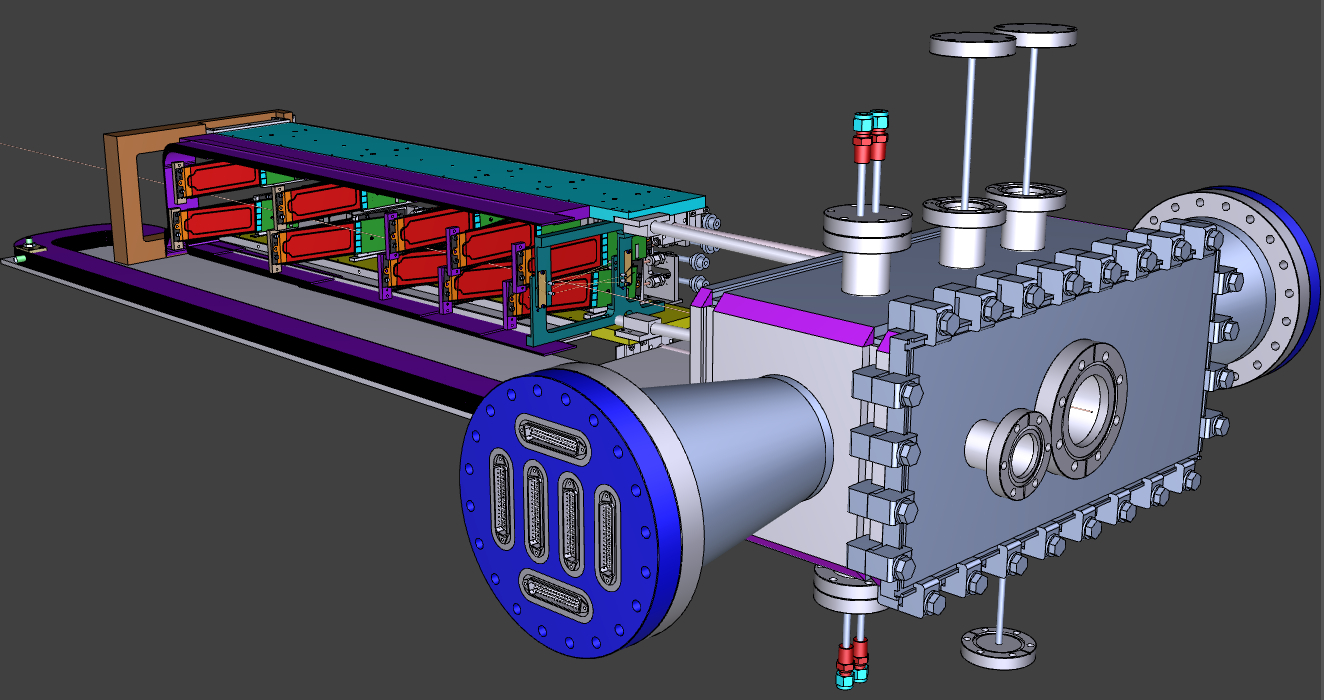}
\caption{\small A rendering of the SVT showing the modules on their support plates held by the 
hinged C-support on the left and the motion levers on the right. The sensors are shown in red and the 
hybrid readout boards in green. The beam enters from the right through a vacuum box with flanges 
for services. 
\label{fig:tracker_model}}
\end{figure}
\end{center}

\begin{figure}[htp]
\begin{center}
    \includegraphics[width=7cm]{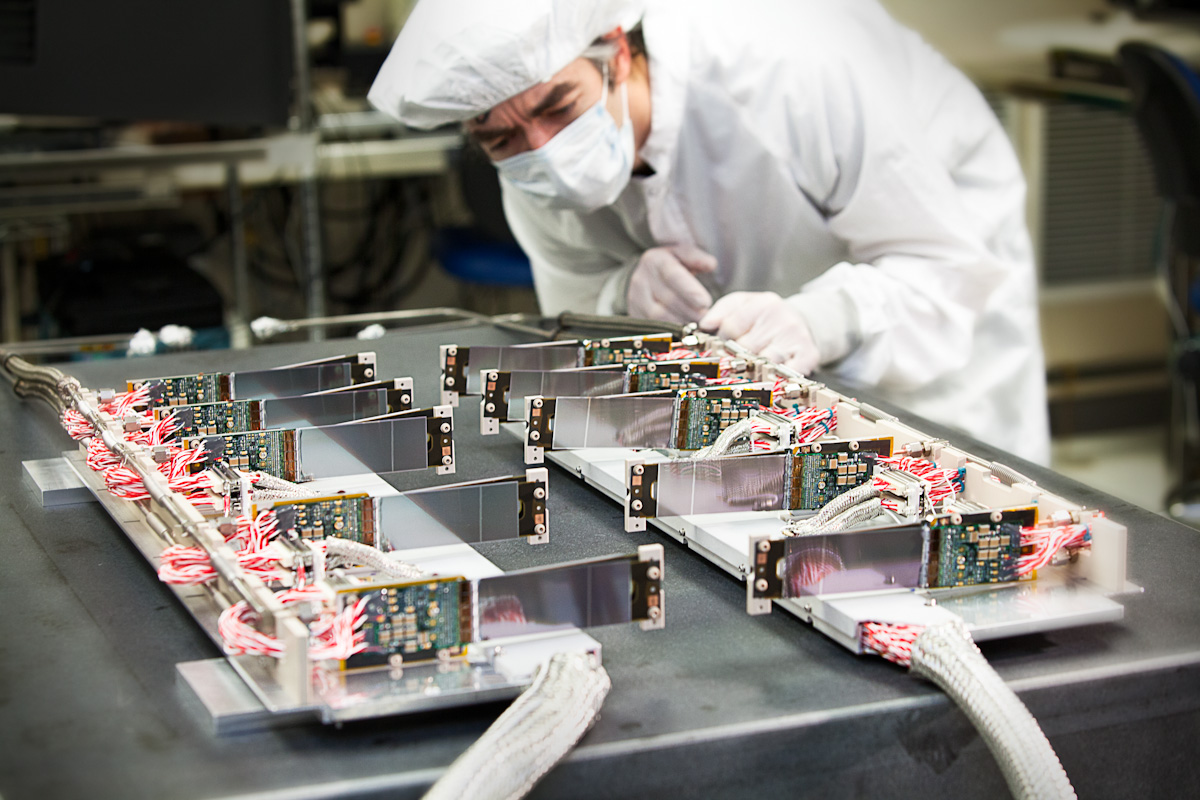}
\caption{\small Both halves of the HPS Test Run SVT after final assembly at SLAC.  The cooling manifolds and 
integrated cable runs are clearly seen.
\label{fig:tracker_halves}}
\end{center}
\end{figure}

The layout of the SVT is summarized in Table~\ref{tab:trk} and rendered in Figure~\ref{fig:tracker_model}. 
Each of the layers is comprised of a pair of closely-spaced silicon microstrip sensors mounted 
back-to-back to form a module. A 100~mrad stereo angle is used in the first three layers to provide 
higher-resolution 3D space points for vertexing.  Using 50~mrad in the last two layers breaks the tracking 
degeneracy of having five identical layers and minimizes fakes from ghost hits to improve pattern 
recognition. Altogether, the SVT has 20 sensors for a total of 12780 readout channels. 
\begin{center}
\begin{table}[ht]
{\footnotesize
\begin{tabular}{lccccc}   
\hline \hline 
    Layer & 1 & 2 & 3 & 4 & 5 \\      
\hline
    $z$ from target (cm)  & 10 & 20 & 30 & 50 & 70  \\ 
    Stereo angle (mrad)  & 100 & 100 & 100 & 50 & 50 \\ 
    Bend res. ($\mu$m)  & $\approx$60 & $\approx$60 & $\approx$60 & $\approx$120 & $\approx$120  \\ 
    Non-bend res. ($\mu$m)  & $\approx$6 & $\approx$6 & $\approx$6 & $\approx$6 & $\approx$6  \\ 
    \# of sensors  & 4 & 4 & 4 & 4 & 4  \\ 
    Dead zone (mm) & $\pm1.5$  & $\pm3.0$  & $\pm4.5$  & $\pm7.5$  & $\pm10.5$  \\ 
    Power cons. (W) & 6.9 & 6.9 & 6.9 & 6.9 & 6.9 \\
\hline \hline
\end{tabular}
\caption{\small Layout of the SVT.
\label{tab:trk}}
}
\end{table}
\end{center}
The SVT is built in two separate halves that are mirror reflections of one another about the plane of 
the nominal electron beam.  Each half consists of five modules mounted on a support plate that 
provides services to the modules and allows them to be moved as a group relative to the dead zone. 
The two halves of the tracker are connected to hinges mounted on a C-shaped support just beyond 
the last layer that defines the nominal spacing between the upper and lower halves of the tracker.  A 
shaft attached to each support plate in front of layer one extends upstream and connects to a linear shift 
that transfers motion into the vacuum box through bellows to open and close the two halves around 
the dead zone. The C-support is mounted to an aluminum baseplate that defines the position of the 
SVT with respect to the vacuum chamber. Figure~\ref{fig:tracker_halves} shows a photograph of both 
completed detector halves prior to final assembly. 

\subsection{Components}
The sensors for the SVT are $p+$-on-$n$, single-sided, AC-coupled, polysilicon-biased microstrip 
sensors fabricated on $<$100$>$ silicon and have 30 (60)~$\mu$m sense (readout) pitch over their 
$4\times10$~cm$^2$ surface. This sensor technology was selected to match the requirement of 
$<1$\% $X_0$ per layer, single-hit resolution better than 50~$\mu$m and tolerance of a radiation 
dose of approximately $1.5\times10^{14}$~\fluenceunit{} for a six month run. The sensors, produced by 
Hamamatsu Photonics Corporation, were originally meant for the cancelled 
Run 2b upgrade of the D\O~experiment~\cite{Denisov:2001aa} which satisfied the requirement that 
the technology must be mature and available within the time and budget constraints.

Despite having only small spots with very high occupancy (up to 4~MHz/mm$^2$) closest to the primary 
beam, the rates are still high and lowering the peak occupancy to 
approximately 1\% for tracking requires a trigger window and hit time tagging of roughly 8~ns. The 
ECal readout and trigger described in Sec.~\ref{sec:fadc} can achieve such resolution. To reach this 
performance the sensors for the SVT are readout by the APV25 ASIC developed for the CMS 
experiment at CERN~\cite{French:2001xb}. The APV25 ASIC can capture successive samples of the shaper output 
in groups of three at a sampling rate of approximately 40~MHz.  By fitting the known 
$CR$-$RC$ shaping curve to these samples, the initial time of the hit can be determined to a precision 
of 2~ns for S/N$\approx$25~\cite{Friedl:2009zz}.  For electron beam running, six-sample readout and 
the shortest possible shaping time (35~ns) are used to best distinguish hits that overlap in time.
The APV25 ASICs are hosted on simple FR4 hybrid readout boards outside the tracking volume with a 
short twisted-pair pigtail cable to provide power, configuration, and signal readout. Along with a 
single sensor, these are glued to a polyimide-laminated carbon fiber composite backing making 
up a half-module. A window is machined in the carbon fiber leaving only a frame around the periphery 
of the silicon to minimize material. A 50~$\mu$m sheet of polyimide is laminated to the surface of the 
carbon fiber with 1~mm overhang at all openings to ensure good isolation between the back side of the 
sensor, carrying high-voltage bias, and the carbon fiber which is held near ground. 

\begin{figure}[htp]
	\begin{center}
   	 \includegraphics[width=7cm]{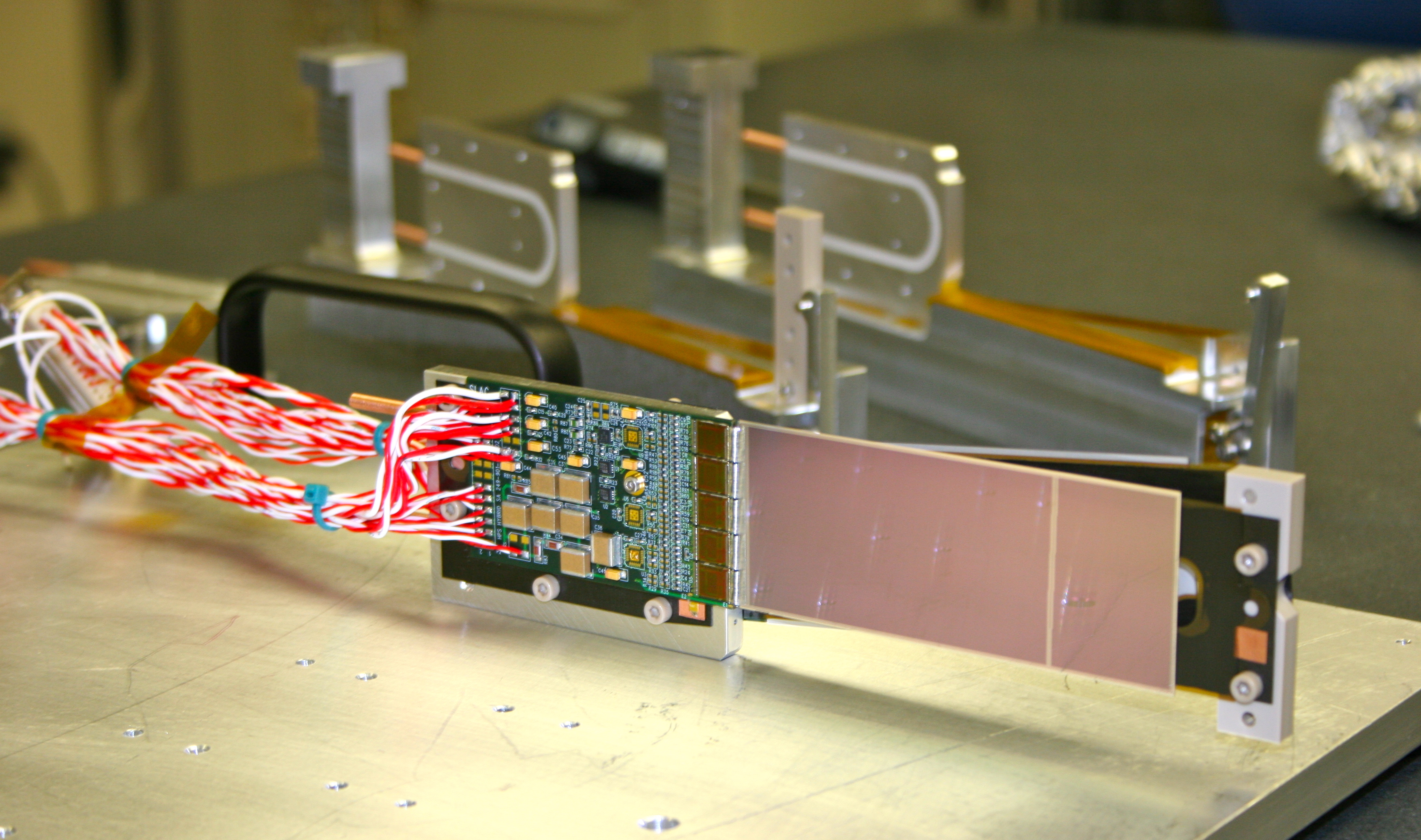}
	\caption{\small A prototype module assembly (foreground) with the 50~mrad (left) and 100~mrad (right) module assembly fixtures in the background.  A pair of cooling blocks and a spacer block can be seen on the fixtures.
\label{fig:tracker_module}}
\end{center}
\end{figure}

The sensor modules for the SVT consist of a pair of identical half-modules, sandwiched back-to-back 
around an aluminum cooling block at one end and a similar PEEK spacer block at the other. 
Figure~\ref{fig:tracker_module} shows a single module after assembly.
The cooling block provides the primary mechanical support for the module as well as cooling via copper 
tubes pressed into grooves in the plates. The spacer block defines the spacing between the sensors at 
the far end of the module, stiffens the module structure, and improves the stability of the sensor 
alignment.  The average support material in the tracking volume is approximately 0.06\% $X_{0}$ per 
double-sided module for a total material budget of 0.7\% per layer.

The total SVT power consumption budget of about 50~W is removed by a water/glycol mixture 
circulated through a flexible manifold attached to the copper tubes in the cooling blocks. During the 
HPS Test Run the sensors were operated at around $23^{\circ}$~C. The power consumption is 
dominated by five 
APV25 ASICs on each hybrid board consuming approximately 2~W, the radiant heat load is less than 
0.5~W per sensor and leakage current is only significant in a small spot after irradiation.

\subsection{Production, Assembly and Shipping}

Hybrids with APV25 ASICs underwent quick qualification testing and each half-module was run at low 
temperature ($\approx5^{\circ}$ C) and fully characterized for pedestals, gains, noise and time 
response after assembly.  Of 29 half-modules built, 28 passed qualification testing, leaving eight spare 
modules after completion of the SVT. Only sensors capable of 1000~V bias voltage without breakdown 
were used.  Full-module assembly 
and mechanical surveys were performed at SLAC before final assembly, testing and shipping of the 
SVT to JLab. A custom shipping container with nested crates and redundant isolation for shock and 
vibration was built in order to safely send the partly assembled SVT to JLab. At JLab, the entire SVT was 
integrated with the full data acquisition and the power supplies before moving the module-loaded support plates to 
Hall B for final mechanical assembly and installation inside of the vacuum chamber.

\subsection{Alignment}
The SVT was aligned using a combination of optical, laser and touch probe surveys at SLAC and JLab. 
The optical survey of individual modules with a precision of a few $\mu$m was combined with a 
touch-probe survey of the overall SVT support structure, with 25-100~$\mu$m precision, to locate the 
silicon sensor layers with respect to the support plates and the mechanical survey balls on the base 
plate. After full assembly and installation of the SVT at JLab, a mechanical survey of the SVT base plate 
position inside the pair spectrometer vacuum chamber is used to determine the global position of the 
SVT with respect to the CEBAF beam line. The resulting survey-based alignment has the position of the 
silicon sensors correct to within a few hundred microns measured from tracks in the HPS Test Run 
data. A more sophisticated global track-based alignment technique to reach final alignment precision 
well below 50~$\mu$m is being developed.


\section{Electromagnetic Calorimeter}
\label{sec:ecal}

The electromagnetic calorimeter (ECal), installed downstream of the PS dipole magnet, performs two 
essential functions for the experiment: it provides a trigger signal to select what events to read out from 
the detector sub-systems and is used to identify electrons and positrons. 
The technology and design choices are largely driven by the need for a compact forward design 
covering the SVT \Aprime{} 
acceptance and the ability to measure the energy and positions of electrons and positrons with energy between
0.5 and 6.6~GeV. It needs granularity and signal 
readout speed to handle 1~MHz/cm$^{2}$ of electromagnetic background as well as good radiation hardness. 
Even modest energy resolution is adequate for triggering. HPS requires better 
energy resolution, $\sigma(E)/E<5\%/\sqrt{E}$, so that the ECal energy measurement can be used in 
combination with that from the SVT to improve the overall momentum resolution.

The PbWO$_{4}$ crystal inner calorimeter of the CLAS detector~\cite{clas_ecal,clas_ecal2,clas_thesis}, in 
operation since 2005 in Hall B, meets all the requirements set by HPS. The modules from this calorimeter have been subsequently 
repurposed for HPS.

\subsection{Components}

\begin{figure}[]
\begin{center}
\includegraphics[width=7cm]{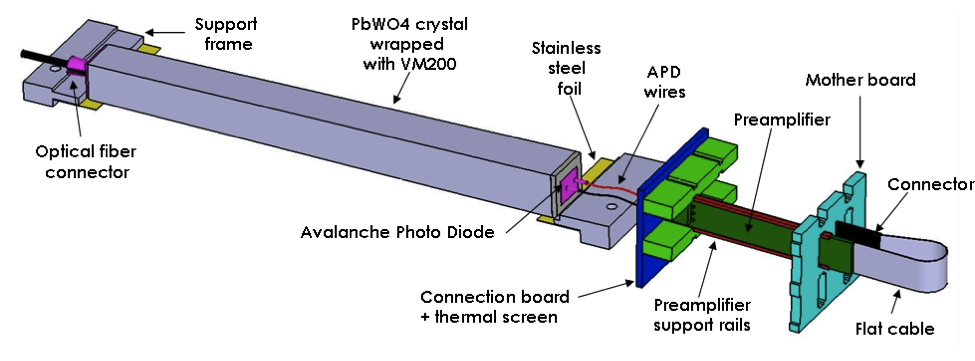}
\caption{\small A schematic view of an ECal module.
\label{fig:ecal-module}}
\end{center}
\end{figure}

The ECal module shown in Figure~\ref{fig:ecal-module} is based on a tapered 160~mm long 
PbWO$_{4}$ crystal with a $13.3\times13.3$~mm$^2$ ($16\times16$~mm$^2$) front 
(rear) face wrapped in VM2000 multilayer polymer mirror film. 
The scintillation light yield, approximately $120$ photons/MeV, is read out by a 
5$\times$5~mm$^2$ Hamamatsu S8664-55 Avalanche Photodiode (APD) with 75\% quantum 
efficiency glued to the rear face surface using MeltMount 1.7 thermal plastic adhesive. This 
results in about eight photoelectrons/MeV which needs to be amplified before being fed into the JLab 
Flash ADC~\cite{fadc} (FADC) board for digitization and 
processing. The maximum energy deposited in a crystal is expected to be 4.2~GeV which needs to 
match the input range of the FADC. The relatively low gain of the APD ($\sim$200) was compensated with 
custom-made preamplifier boards, that provide further amplification to match the 2~V dynamic range of the 
FADC. The ADC has 12-bit resolution. Gains are adjusted to give about 1~ADC count/MeV.
This dynamic range is adequate to measure the 10~MeV noise level as well as the 
the maximum energy expected in a single crystal, about 4~GeV.

\subsection{Layout}
\begin{figure}[]
\begin{center}
\includegraphics[width=0.45\textwidth]{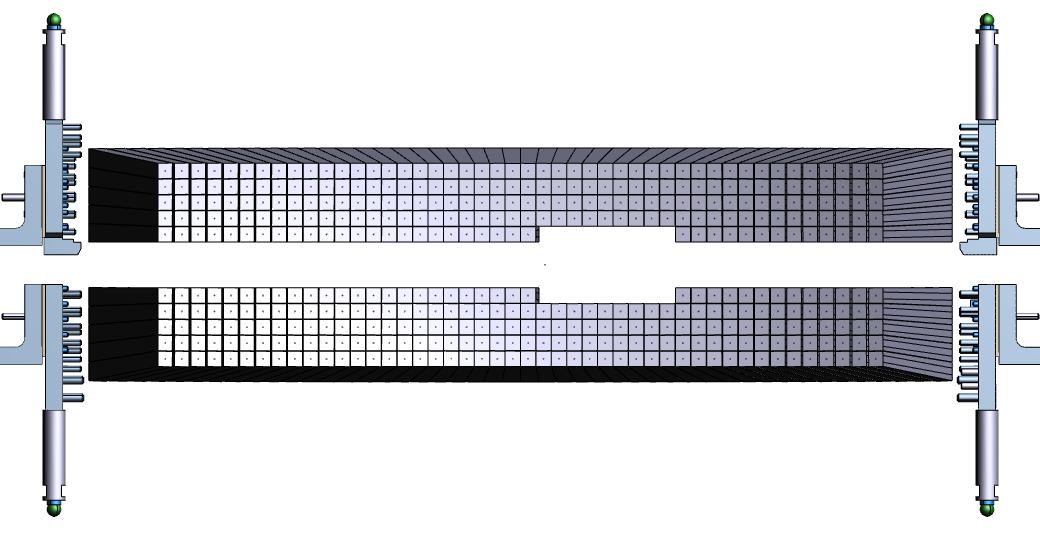}
\caption{\small Rendered layout view of the ECal looking downstream.
\label{fig:ecal}}
\end{center}
\end{figure}

Similar to the SVT, the ECal is built in two separate halves that are mirror reflections of one another 
about the plane of the nominal electron beam to avoid interfering with the 15~mrad ``dead zone''. 
As shown in Figure~\ref{fig:ecal}, the 221 modules in each half, supported by aluminum support frames, 
are arranged in rectangular formation with five layers and 46 crystals/layer except for the layer closest to 
the beam where nine modules were removed to allow a larger opening for the outgoing electron and 
photon beams. Each half was enclosed in a temperature controlled box ($<1^{\circ}$~F stability and 
$<4^{\circ}$~F uniformity) to stabilize the crystal light yield and the operation of the APDs and its 
preamplifiers. 
Four printed circuit boards mounted on the backplane penetrated the enclosure and were used to 
supply the $\pm 5$~V operating voltage for the preamplifiers, 400~V bias voltage to the APDs, and to 
read out signals from the APDs. Each half of the ECal was divided into 12 bias voltage groups with a 
gain uniformity of about 20\%. 

During the HPS Test Run, both halves were held in place by four vertical bars attached to a rail above, 
placing the front face of the crystals 147~cm from the upstream edge of the magnet, with a 
8.7~cm gap between the innermost edge of the crystals in the two halves.

\subsection{Signal readout}
\label{sec:fadc}
After a 2:1 signal splitter, $1/3$ of an amplified APD signal was fed to a 
single channel of a FADC board. $2/3$ of the signal was sent to a 
discriminator module and then to a TDC for a timing measurement. 
The FADC boards are high speed VXS modules digitizing up to 16 APD signals 
at 250~MHz and storing samples in 8~$\mu$s deep pipelines with 12-bit resolution. 
When a trigger is received, the part 
of the pipeline from five samples before and 30 after the signal which crossed a programmable 
threshold (for the HPS Test Run this was set to $\approx70$~MeV) are 
summed and stored in a 17-bit register for readout. In addition a 4~ns resolution timestamp of the 
threshold crossing is reported in the readout for each pulse.
This scheme significantly compresses the data output of the FADC. During offline data analysis, a 
calibrated pedestal value is subtracted to obtain the actual summed energy.
Two 20-slot VXS crates with 14 (13) FADC boards were employed in the HPS Test Run to read out the 
top (bottom) half of the ECal.


\section{Trigger and data acquisition}
\label{sec:triggerdaq}

The data acquisition (DAQ) system handles acquisition of data from the ECal and SVT sub-detectors  with 
two DAQ architectures. The SVT DAQ is based on Advanced Telecom Communications Architecture 
(ATCA) hardware while the ECal uses VMEbus Switched Serial (VXS) based hardware. Data from the 
sub-detectors are only read out when a trigger signal from the trigger system is received.

\subsection{Trigger system}
\label{sec:trigger}
\begin{figure}[b]
\begin{center}
 \includegraphics[width=8cm]{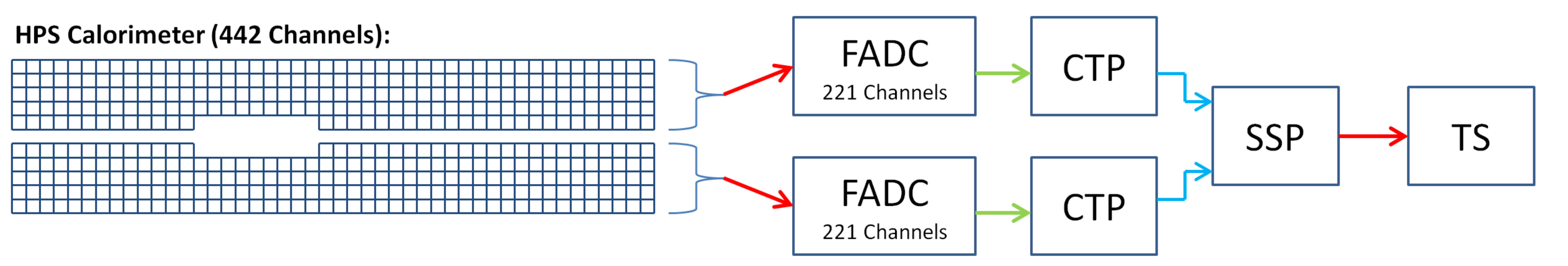}
\caption{\small Block diagram of the ECAL trigger system consisting of the FADC that samples and digitizes 
signals for each detector channel and sends them for cluster finding in the CTP. The CTP clusters are 
sent to the SSP where the final trigger decision is taken based on pairs of clusters in both halves of the 
ECal. The decision is sent to the Trigger Supervisor (TS) that generates the necessary signals to read out 
the sub-detectors.
 \label{fig:hps_trigger_cal}}
\end{center}
 \end{figure}

The trigger system is designed to select time coincidences of electromagnetic clusters in the top and bottom 
halves of the ECal which meets kinematic conditions satisfied by \Aprime{} decays and minimize backgrounds.
The trigger system needs to be essentially dead-time free, handle rates up to 50~kHz, and supply a trigger 
signal which jitters $<8$~ns from the actual event time in order to minimize backgrounds from out-of-time hits in the SVT. 
Figure~\ref{fig:hps_trigger_cal} shows a schematic overview of each stage of the system. 
Each channel on the FADC board has an independent data path to send 5-bit pulse energy and 3-bit 
pulse arrival time information every 32~ns to a Crate Trigger Processor board (CTP), which is in the same 
crate. The 3-bit pulse arrival time allows the trigger to know the pulse timing at 4~ns resolution. 
Contrary to the readout path 
described in Sec.~\ref{sec:fadc}, this energy is a pedestal-subtracted time-over-threshold sum with 
programmable offsets and minimum threshold discriminator for each channel. With input from all 
FADC channels, i.e. one half of the ECal, the CTP performs cluster finding and calculates cluster 
energy and timing information. The 3x3 fixed-window, highly parallel, FPGA-based cluster 
algorithm simultaneously searches for up to 125 clusters with energy sum larger than 
the programmable energy threshold set to about 270~MeV. This high threshold didn't hurt the trigger rate 
rate studies for the HPS Test run since only clusters with high energies were studied but for HPS this 
threshold will need to be lower.  Crystals in the 
fixed-window are included in the sum if the leading edge of the pulse occurred within a 32~ns time 
window to take into account clock skew and jitter throughout the system.
The CTP only accepts clusters with the locally highest energy 3x3 window to deal with overlapping and 
very large clusters. The Sub-System Processor board (SSP) receives the clusters from the top and bottom half CTP 
at a maximum rate of 250~MHz and searches for pairs of clusters in an 8~ns wide coincidence window. The 
SSP sends triggers to the trigger supervisor (TS), which generates all the necessary signals and 
controls the entire DAQ system readout through the trigger interface units installed in every crate that 
participate in the readout process.

The trigger system is free-running and driven by the 250~MHz global clock and has essentially zero 
dead time at the occupancies expected for HPS. The trigger supervisor can apply dead time if 
necessary, for example on a `busy' or `full' condition from the front-end electronics. The system is 
designed to handle trigger rates above 50~kHz and has a latency set to $\approx 3~\mu$s to match 
that required by the SVT APV25 ASIC. 

During most of the HPS Test Run, the trigger system 
required only a single cluster in either the top or bottom ECal halves resulting in rates below 2~kHz. However, 
the trigger system was tested to rates above 100~kHz by lowering thresholds.

\subsection{SVT data acquisition}
\label{sec:svt_daq}
The purpose of the SVT DAQ is to support the continuous 40~MHz readout and processing of signals from 
each of the 20 silicon strip sensors of the SVT. The data for each strip channel, six samples of the 
signal, needs to be transferred to the JLab DAQ for those events selected by the trigger at rates 
up to 50~kHz and with data transfer rates up to 100~MB/s.

\begin{figure}[]
\begin{center}
\includegraphics[width=6cm]{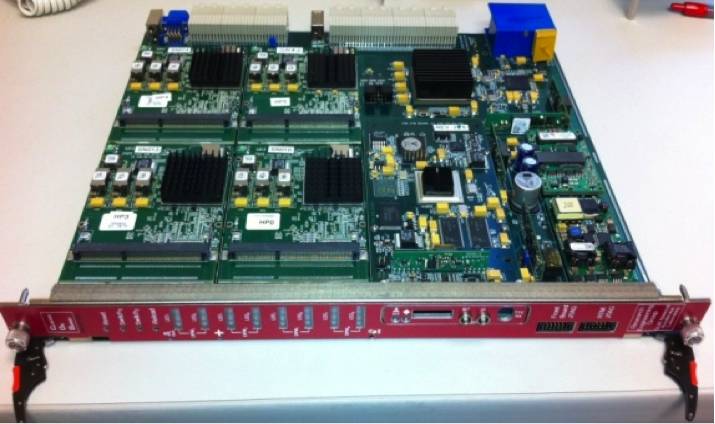}
\caption{\small The SVT DAQ COB board with four data processing daughter cards (DPMs) visible on the left side.
\label{fig:cob}}
\end{center}
\end{figure}

\begin{figure}[]
\begin{center}
\includegraphics[width=8cm]{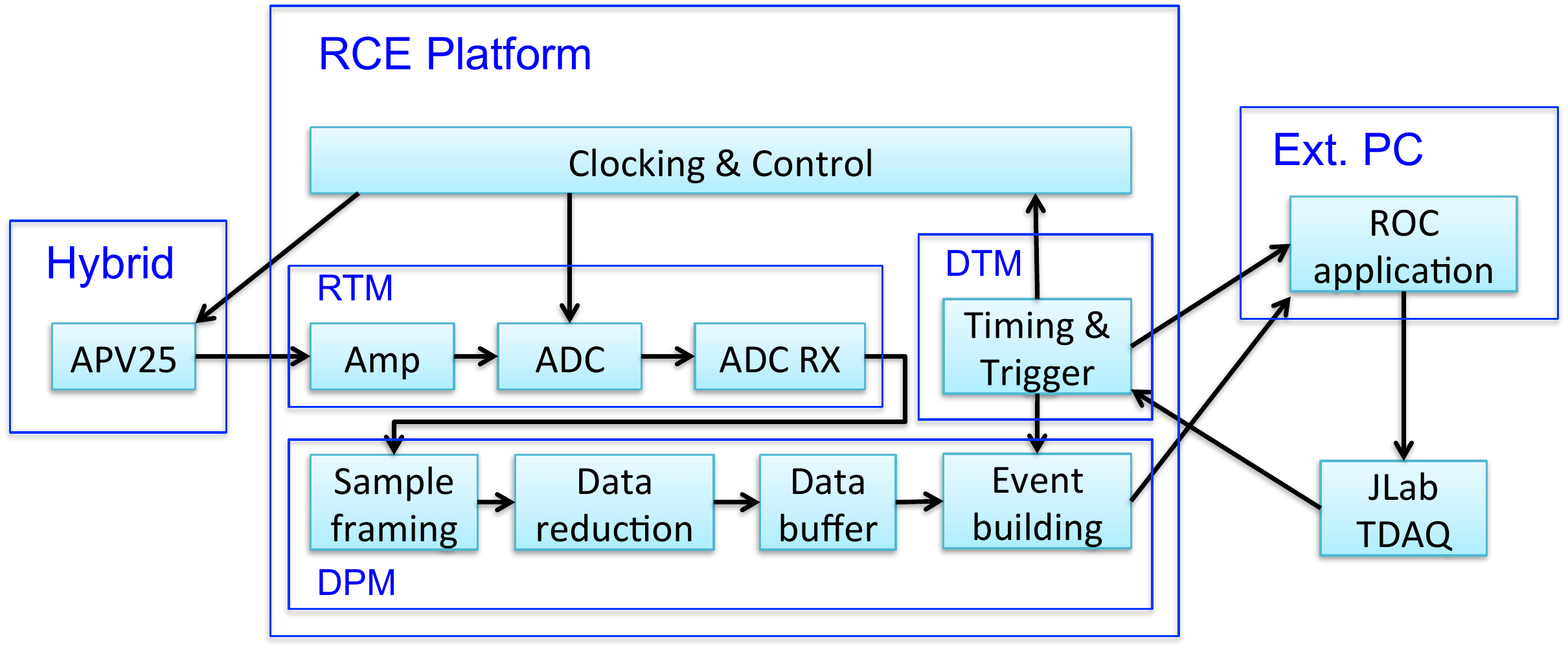}
\caption{\small Block diagram overview of the SVT DAQ.
\label{fig:svtdaq}}
\end{center}
\end{figure}

The SVT DAQ is based on the Reconfigurable Cluster Element (RCE) and cluster 
interconnect concept developed at SLAC as generic building blocks for DAQ systems. 
The RCE is a generic computational building block, housed on a separate daughter card called 
Data Processing Module (DPM), that is realized on an ATCA front board called the Cluster On Board 
(COB), as shown in Figure~\ref{fig:cob}.
The first generation RCE used in the HPS Test Run consisted of a Virtex~5 FPGA with 1~GB of DDR3 RAM. 
A schematic overview of the system is shown in Figure~\ref{fig:svtdaq}. 
The analog outputs of up to 12 SVT half-modules (60 APV25 ASICs) are digitized on the 
Rear-Transition-Module (RTM), a custom board on the back side of the ATCA crate, interfacing the 
HPS-specific readout to the generic DAQ components on the COB. A preamplifier converts the APV25 ASIC 
differential current output to a different voltage output scaled to the sensitive range of a 14-bit ADC 
operating at the system clock of 41.667~MHz. The RTM is organized 
into four sections with each section supporting three SVT half-module hybrids (15 APV25 ASICs). The 
RTM also includes a 4-channel fiber-optic module and supporting logic which is used to interface 
to the JLab trigger system supervisor. Each section of the RTM is input to a DPM which apply 
thresholds for data reduction and organizes the sample data into UDP datagrams. The DPM also hosts 
an I$^{2}$C controller used to configure and monitor the APV25 ASICs. A single ATCA crate with two 
COB cards was used, one supporting four DPMs and one supporting three DPMs and one DPM that is 
configured as the trigger and data transmission module. The two COB cards and their DPMs are 
interconnected with a 10~Gb/s switch card~\cite{Larsen:2011zb} which also hosts two 1~Gb/s Ethernet 
interfaces to the external SVT DAQ PC.  

The external PC supports three network interfaces: two standard 1~Gb/s Ethernet and one custom 
low-latency data reception card. The first is used for slow control and monitoring of the eight 
DPM modules and the second serves as the interface to the JLAB data acquisition system. The third 
custom low-latency network interface is used to receive data from the ATCA crate and supports a low 
latency, reliable TTL trigger acknowledge interface to the trigger DPM. This PC hosts the SVT control 
and monitoring software as well as the readout controller application used to interface with the 
JLab DAQ.

In order to minimize cable length for the analog APV25 ASIC output signal the ATCA crate was located 
approximately 1~m from the beam line, next to the cable vacuum feed-throughs.   
Before shielding with lead-blankets and borated polyethylene was arranged, we observed two failures of normally reliable ATCA 
crate power supplies, time-correlated to beam instabilities. 

Although trigger rates during the HPS Test Run were significantly lower, this system was tested at 
trigger rates up to 20~kHz and 50~MB/s.  With optimized event blocking and improved Ethernet 
bandwidth, together with utilizing the overlapping readout and trigger functionality of the APV25 ASIC, the 
system is capable of being read out at 50~kHz trigger rate.

\subsection{General data acquisition and online computing}
\label{sec:daq}
Every crate participating in the readout process contains a Readout Controller (ROC) that 
collects digitized information, processes it, and sends it on to the event builder. For the ECal, both 
VXS crates run ROC applications in a single blade Intel-based CPU module running CentOS 
Linux OS. For the SVT DAQ, the ROC application runs on the external PC under RHEL. 
The event builder assembles information from the ROCs into a single event which is passed to the 
event recorder that writes it to a RAID5-based data storage system capable of handling up to 
100~MB/s. The event builder and other critical components run on multicore Intel-based multi-CPU 
servers. The DAQ network system is a network router providing 10~Gb/s high-speed connection 
to the JLab computing facility for long-term storage. For the HPS Test Run, both the SVT and ECal ROC had 
a 1~Gb/s link to the network router.


\section{Reconstruction and performance}

While dedicated electron beam was precluded for the HPS Test Run the short dedicated photon beam run 
allowed the study of some of the key performance parameters for HPS and the trigger rates expected during 
electron beam running. This section documents the performance and discusses the implications of these 
results for HPS.

\subsection{SVT performance}

For the duration of the HPS Test Run all SVT modules and APV25 ASICs were configured to their 
nominal operating points~\cite{Jones:1069892} with all sensors reverse-biased at 180~V.  The 
sensors were operated within a temperature range of  $20-24^\circ$C.
Approximately 97\% of the 12,780 SVT channels were found to
be operating normally;  the fraction of dead or noisy channels varied from 2.4\% to 4.7\% throughout 
the HPS Test Run. Most of these losses were due to 2-4 misconfigured APV25 ASICs, a known noisy 
half-module and problems in two particular APV25 ASICs.

\subsubsection{Cluster and hit reconstruction}
Track reconstruction in the SVT puts stringent requirement on the clustering and hit reconstruction. The 
multiple scattering in the tracking material dominates the uncertainty in the track parameter estimation and 
effectively determines the roughly 50~$\mu$m (100~$\mu$m) requirement on the spatial hit resolution in 
the non-bend (bend) plane.  The high 
occupancy due to multiple scattered beam electrons in the target close to the beam requires a hit time 
resolution of 2~ns to efficiently reject out-of-time hits in HPS. Both the hit time, based on a fit to the 
APV25 ASIC pulse shape, and the spatial position reconstruction rely on having $S/N$ around 25 for the 
sensors used in HPS. 

\begin{figure}[]
\begin{center}
\includegraphics[width=7cm]{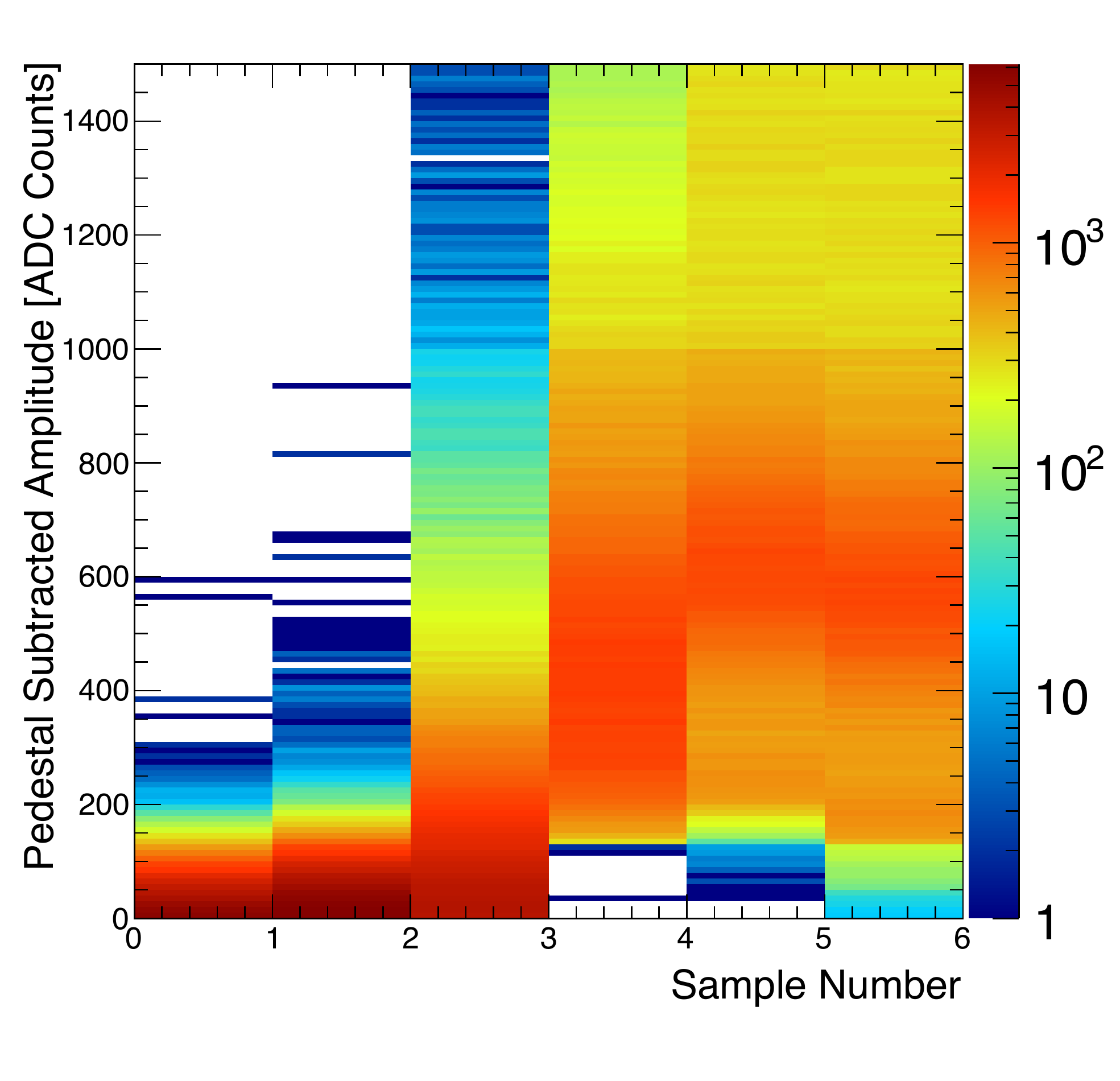}
\caption{\small Accumulation of six pedestal-subtracted samples from individual SVT channels associated 
with hits on tracks.
\label{fig:pulse_shape}}
\end{center}
\end{figure}

\begin{figure}[]
\begin{center}
\includegraphics[width=7cm]{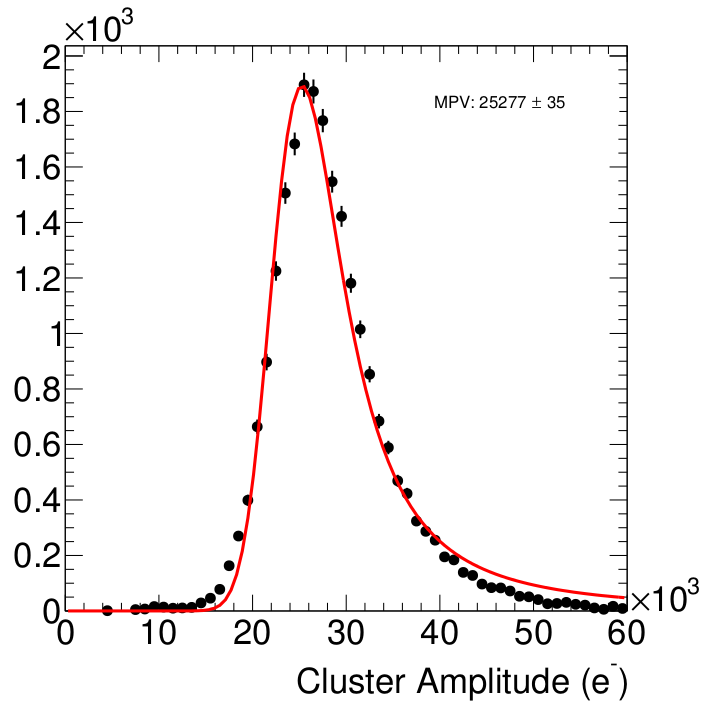}
\caption{\small The cluster charge distribution for hits associated with a track follow the characteristic Landau shape. 
\label{fig:cluster_pulse}}
\end{center}
\end{figure}

\begin{figure}[]
\begin{center}
\includegraphics[width=7cm]{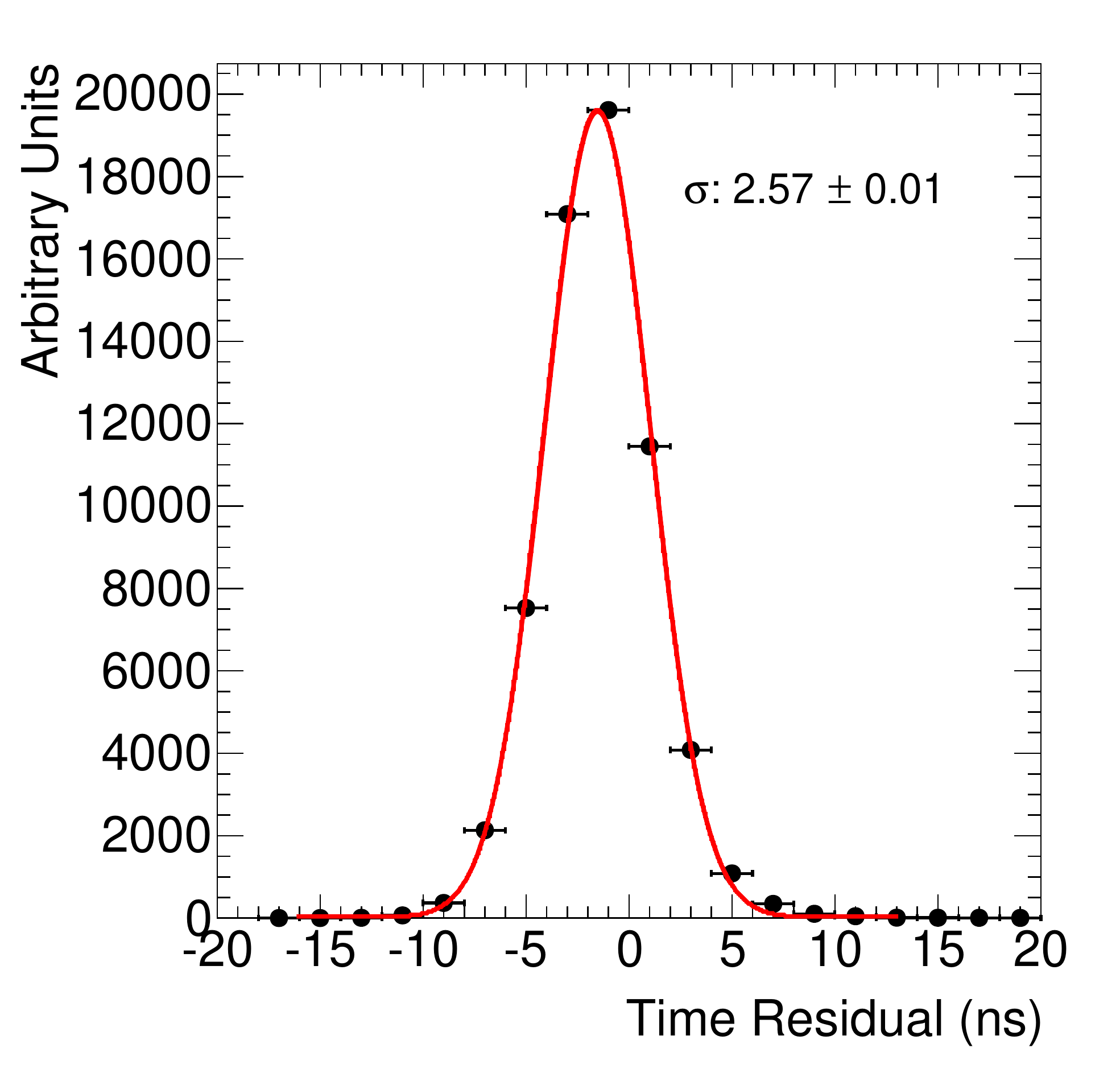}
\caption{\small The residual of individual cluster times with the average of all clusters on the track. 
\label{fig:tracktime}}
\end{center}
\end{figure}

After a trigger is received, the amplitude of every APV25 ASIC is sampled and digitized in the six consecutive time 
bins associated with the trigger time. A data reduction algorithm is applied 
requiring three out of six samples to be above two times the noise level and that the third sample is 
larger than the second or that the fourth sample is larger than the third.
The typical, pedestal subtracted, pulse shape obtained is shown in Figure~\ref{fig:pulse_shape}. 
As the figure demonstrates,  the SVT was well timed-in to the trigger with the rise of the pulse at the 3rd 
sampling point. 
In order to find the time, $t_0$, and amplitude of each hit, the six samples from each channel are fitted 
to an ideal $CR-RC$ function. Note that in the HPS Test Run the APV25 ASICs were operating with a 50~ns 
shaping time. These hits are passed through a simple clustering algorithm which forms clusters by 
grouping adjacent strips with the position of a cluster on the sensor determined by the 
amplitude-weighted mean.
With a linear gain up to $\approx 3$~MIPs, the cluster charge for hits associated with a track follow 
the characteristic Landau shape, see Figure~\ref{fig:cluster_pulse}.  
A noise level between $1.1-1.5\times 10^{3}$ electrons was established through multiple calibration 
runs giving a $S/N$ of $21-25$, in line with the requirement for HPS. 
Radioactive source tests were used to 
provide the absolute charge normalization.
After clustering hits on a sensor, the hit time for each cluster is computed 
as the amplitude-weighted average of the individually fitted $t_0$ on each channel. The 
$t_0$ resolution is studied by comparing the cluster hit time with the average of all cluster hit times on 
the track shown in Figure~\ref{fig:tracktime}. 
After correcting for offsets from each sensor (time-of-flight and clock phase) and accounting for the 
correlation between the $t_0$ and track time,  the extracted $t_0$ resolution is 2.6~ns. This is 
somewhat worse than the approximately 2~ns resolution expected for S/N$=$25 which we attribute to the 
true pulse shape differing from our idealized fit function which will be improved in the 
future~\cite{Friedl2007385}. Reducing 
the APV25 ASIC pulse shaping time to 35~ns will also improve time resolution. 
These results show that HPS can operate with the six sample readout mode of the APV25 ASIC and 
achieve time resolution adequate for pileup rejection during electron running in HPS. 

\begin{figure}[]
\begin{center}
\includegraphics[width=7cm]{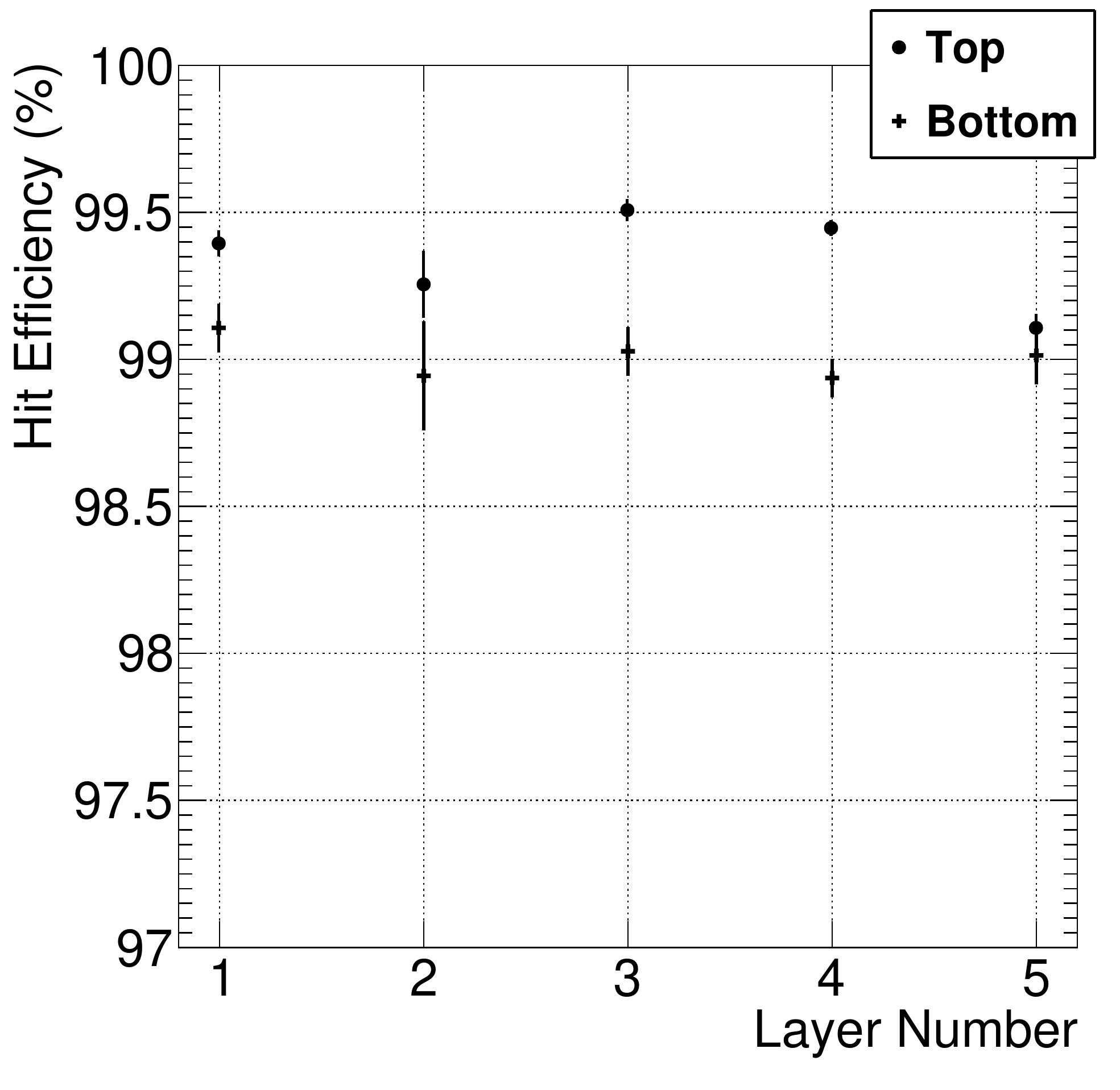}
\caption{\small The hit reconstruction efficiency as a function of detector layer.
\label{fig:hit_efficiency}}
\end{center}
\end{figure}

Good agreement was obtained between observed and simulated occupancies after taking into account 
dead or noisy channels. The hit reconstruction efficiency was estimated by measuring 
the number of good tracks with a hit close to the interpolated position on a given sensor that was excluded 
from the track fit. Tracks which intersect regions with known bad channels or pass very 
close to the edge region were excluded. The hit reconstruction efficiency, see Figure~\ref{fig:hit_efficiency}, 
was measured to be above 98\% and fairly uniform across the SVT. 

The spatial resolution of similar microstrip sensors is well established by test beam data, against which 
the charge deposition model in the simulation is validated.  This resolution can be parameterized as a 
function of the total signal to single-strip noise and the crossing angle of tracks through the sensor.  
The single-hit resolution for charged particles with $S/N>20$, as demonstrated 
here, is relatively constant at approximately 6~$\mu$m for tracks that enter approximately normal to 
the sensors as in HPS. This resolution is significantly better than the requirement for reaching the mass 
and vertex resolutions required for HPS.

\subsubsection{Momentum and vertexing resolution}
\begin{center}
\begin{figure*}[t]
\includegraphics[width=5cm]{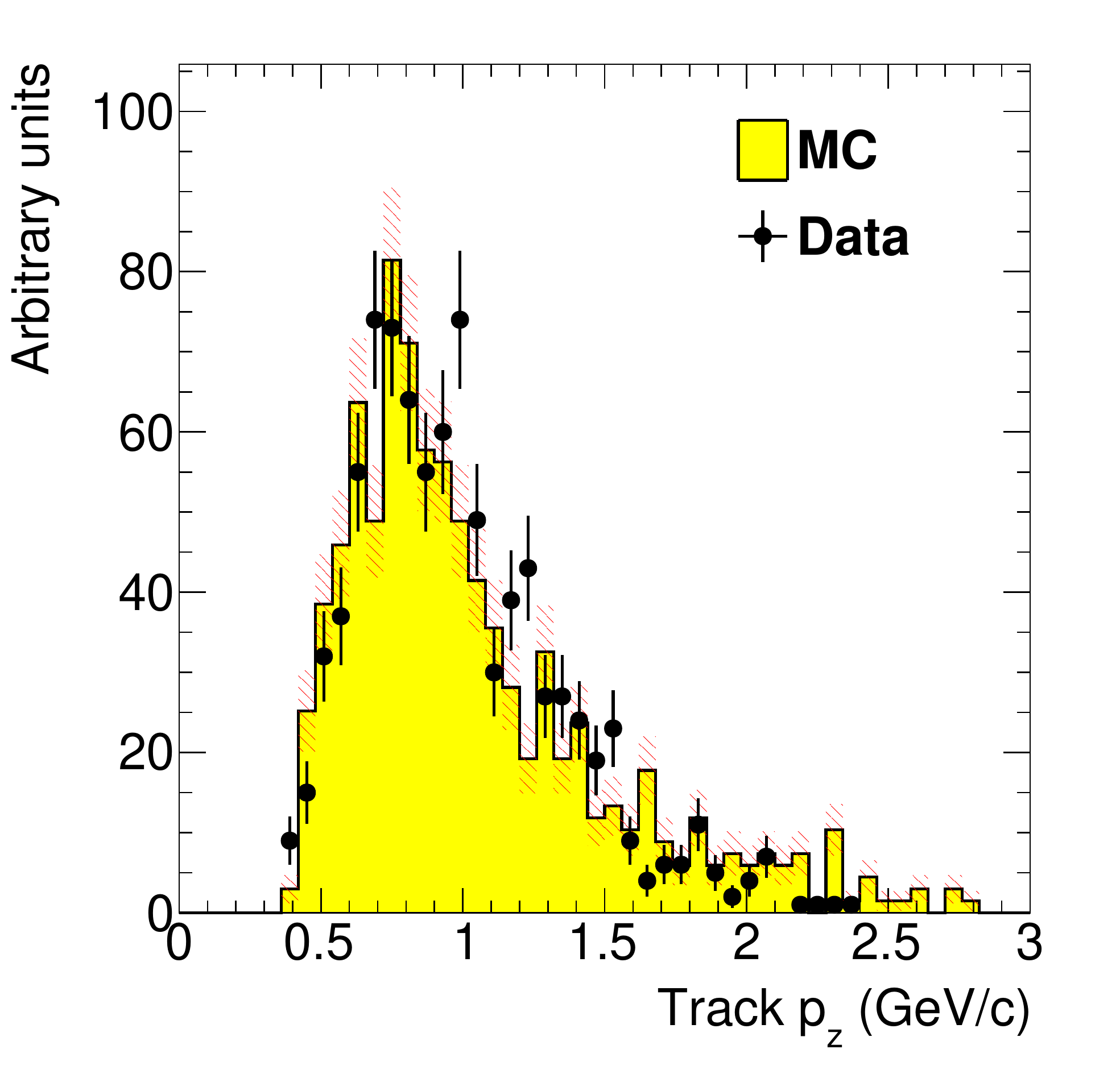}
\includegraphics[width=5cm]{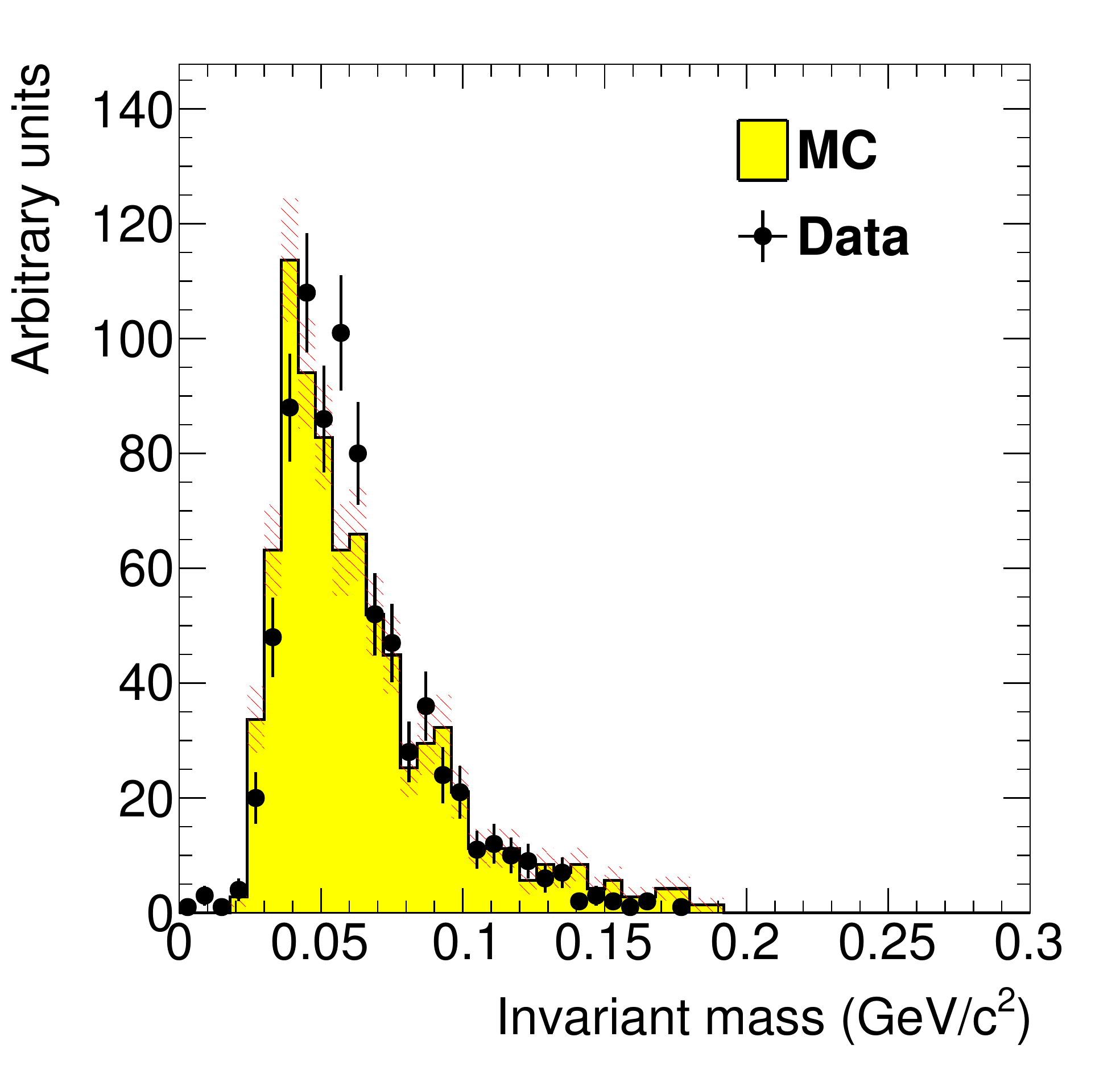}
\includegraphics[width=5cm]{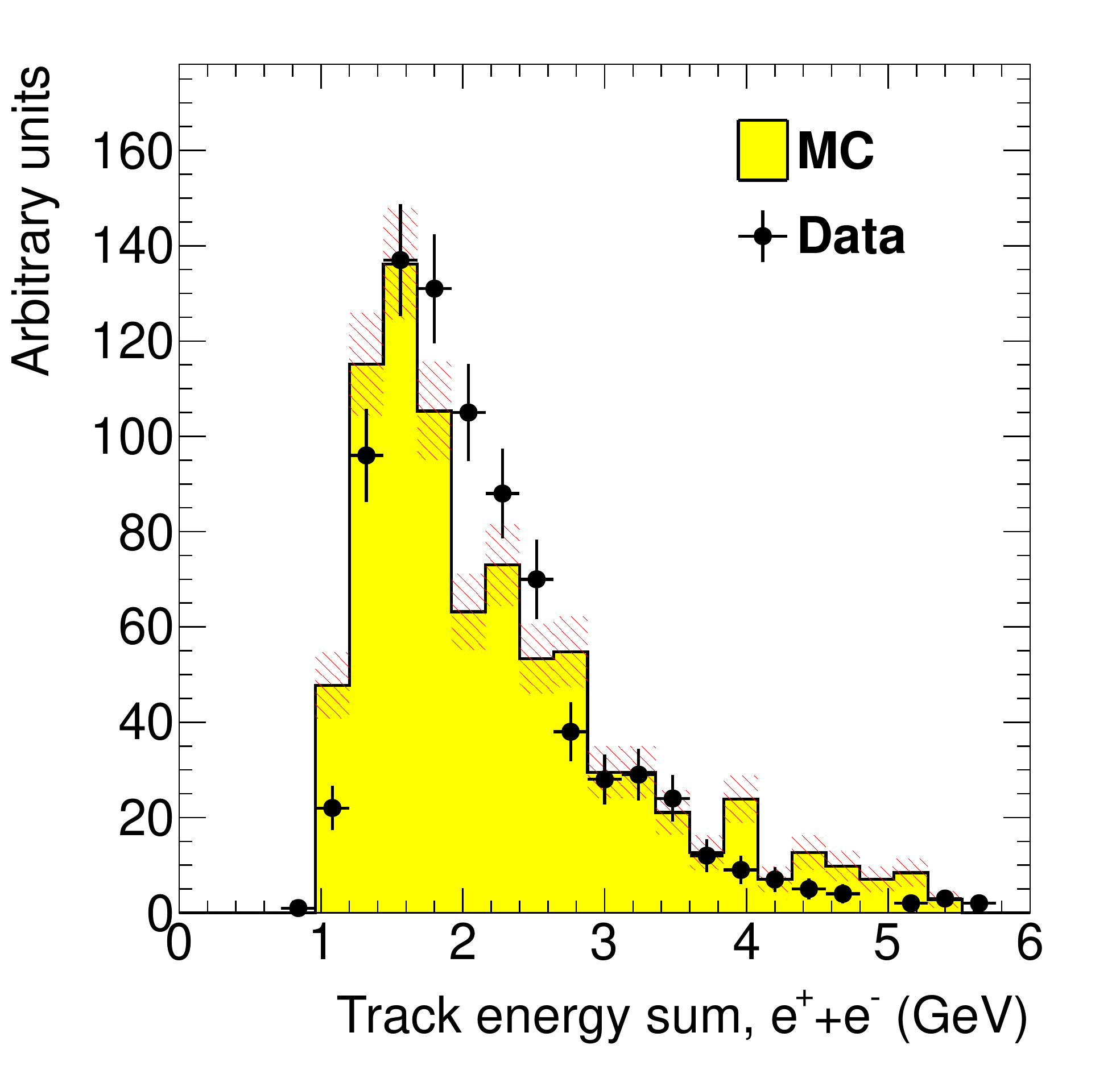}
\caption{\small Kinematic distributions for \ee{} pairs selected by opposite charged tracks in the top and bottom half of the tracker: track momentum in the top half of the SVT (left), invariant mass (middle) and the sum of the track energy for the pair (right). 
\label{fig:pair_kin}}
\end{figure*}
\end{center}

Good track reconstruction performance is crucial to HPS. Simulations show that track momentum resolution 
of 4-5\% is needed to achieve the desired \Aprime{} mass resolution. The precise reconstruction of the 
production vertex to reject prompt QED background requires impact parameter resolutions between 
100-250~$\mu$m for tracks between 0.5-1.7~GeV/c. These key performance parameters where studied 
in the HPS Test Run by selecting \ee{} pairs from photon conversions. Pairs of oppositely charged tracks, 
one in the top and one in the bottom half of the SVT, with momentum larger than 400~MeV/c were selected 
and basic distributions of pair production kinematics were studied.
The kinematics are relatively well reproduced as shown in Figure~\ref{fig:pair_kin}. 

\begin{figure}[]
\begin{center}
	\includegraphics[width=7cm]{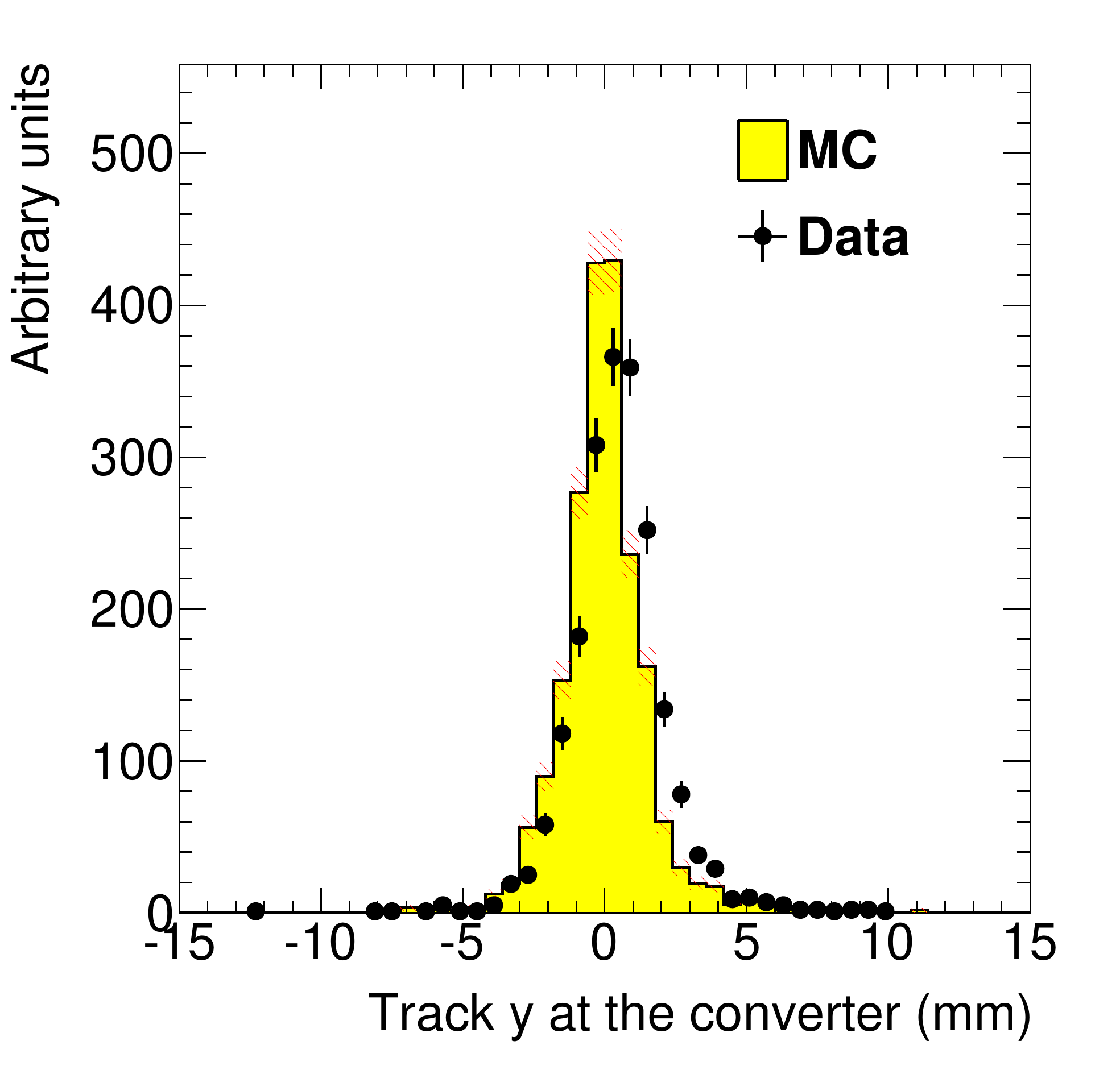}
	\includegraphics[width=7cm]{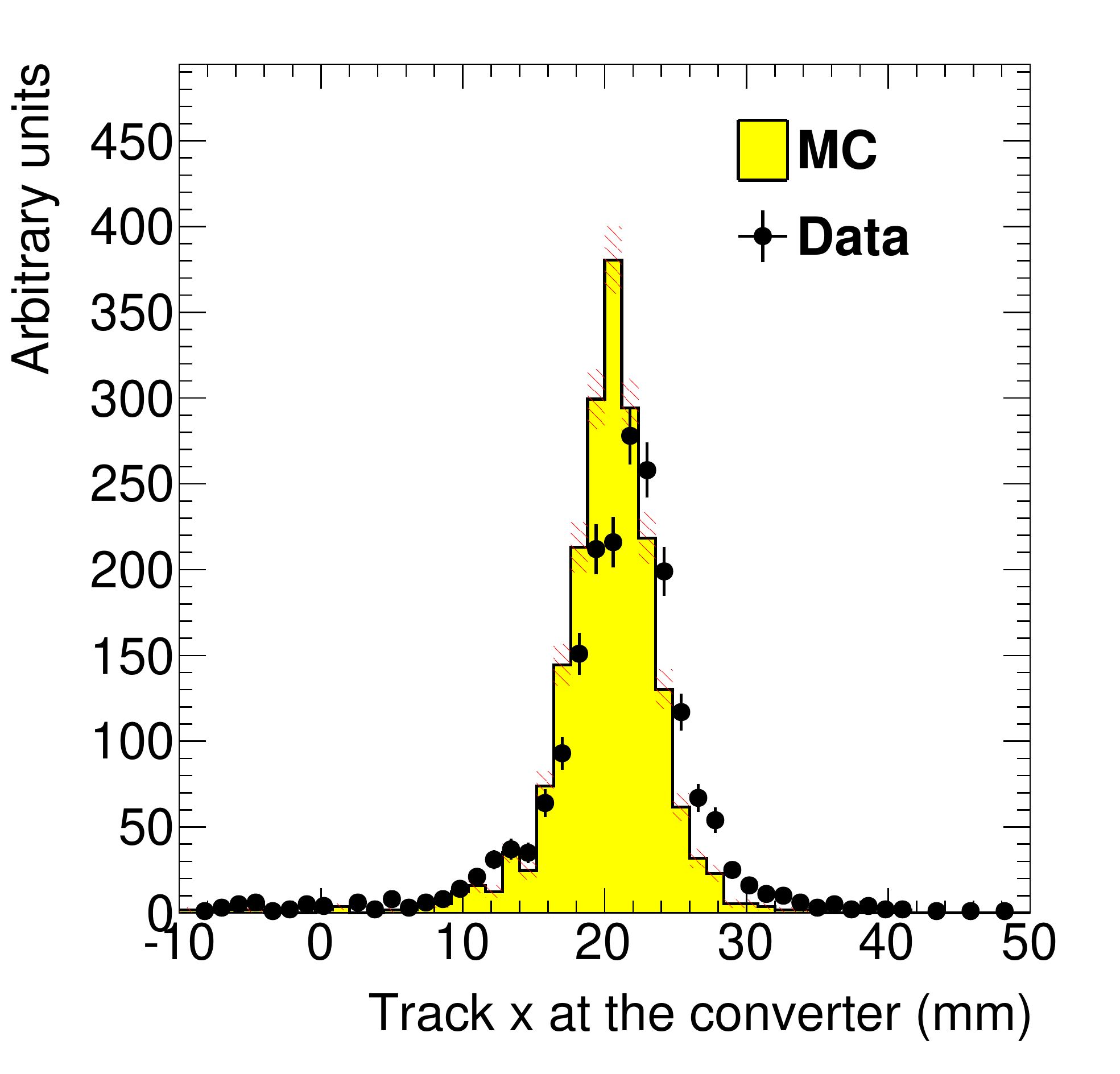}
	\caption{\small Vertical (top) and horizontal (bottom) extrapolated track position at the converter position 
	taking into account the measured fringe field. 	
	\label{fig:impact_param}}
\end{center}
\end{figure}

The expected momentum resolution from simulation is between 4-5\% for tracks in the momentum 
range of the HPS Test Run. By comparing the shapes of the kinematic distributions for data and simulation, we 
estimate an agreement with the nominal scale and resolution to within 10\%.
In the HPS Test Run, as well as in electron running with HPS, the dominant source of uncertainty in the 
tracking and vertexing is multiple Coulomb scattering. For the vertexing performance the foremost 
difference between the HPS Test Run and HPS is that the HPS Test Run target is 67~cm further upstream, 
so tracks must be extrapolated nearly eight times as far as in HPS, giving almost collinear tracks in the 
detector. The increased 
lever arm over which tracks are extrapolated worsens the resolution up to a factor of eight 
(depending on momentum) compared to what is achieved at the nominal electron target position for 
HPS. Figure~\ref{fig:impact_param} shows the horizontal and vertical positions of the extrapolated track 
at the converter position. 
While residual alignments show small shifts, the good agreement  between data and simulated 
events of the widths indicates a good understanding of the material budget and distribution in the SVT.  
Having the dominant contribution to the vertex resolution approximately right demonstrates that the 
Gaussian part of the vertex resolution in HPS, with a target at 10~cm, will be as calculated.

\subsection{ECal performance}
\label{sec:ecal_calibration}
During the HPS Test Run 385 out of 442 modules (87\%) were used in offline reconstruction, 39 modules 
were disabled or not read out (no FADC channel available, no APD bias voltage or masked out due to 
excessive noise) and 18 were masked offline due to noise. 

The integrated pulse of each FADC channel was converted to energy by  
subtracting a pedestal and applying a conversion factor to convert ADC counts to energy. 
The pedestals are measured using special runs where each trigger records 100 samples of signals 
from the APDs with 4~ns between each sample. The pedestals were extracted from the 
part of the window before the actual hit in the calorimeter. Modules with signal above the threshold  
are clustered using a simple algorithm similar to the one 
deployed for the trigger (see Sec.~\ref{sec:trigger}). Due to the high effective crystal readout threshold 
of 73~MeV the average number of crystals in a cluster was only about three and the simple clustering 
algorithm worked well for reconstruction of the detected shower energy. An average noise level of 
approximately 15~MeV per crystal was measured in special pedestal runs. The high crystal noise level 
and effective threshold didn't hurt the trigger rate studies in the HPS Test Run 
as clusters with high energy were used for the analysis. For HPS the noise level and threshold will be 
lowered to improve energy resolution and to allow triggering on cosmics to improve calibration.  

\begin{figure}[]
\begin{center}
	\includegraphics[width=7cm]{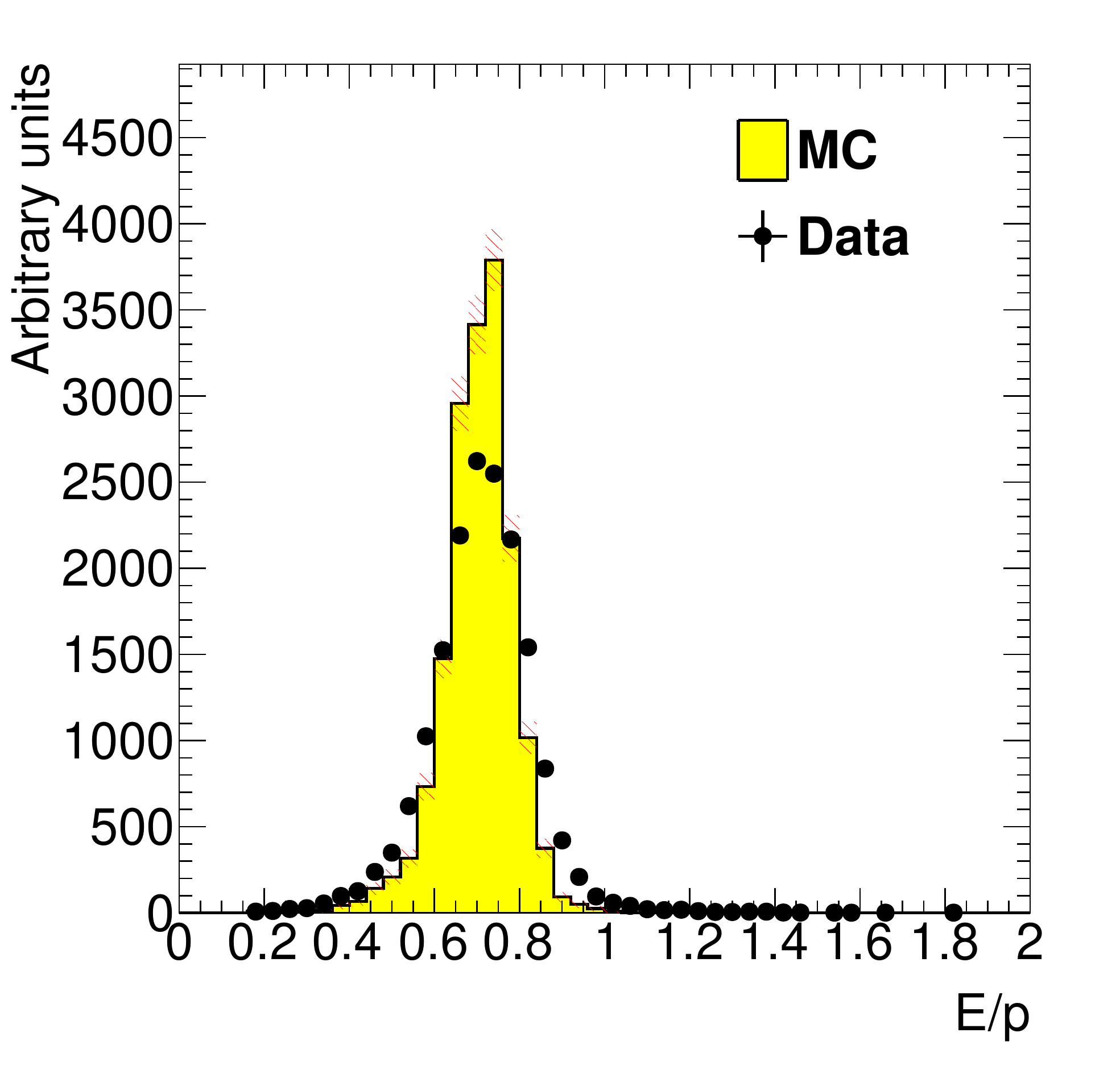}
	\caption{The ECal energy over track momentum ratio ($E$/$p$) comparing data and simulation 
	for single cluster triggers in the top half of the ECal.
	\label{fig:gains}}
\end{center}
\end{figure}

The ratio of the ECal cluster energy $E$ to the momentum $p$ of a matched track in the SVT was used 
to determine the conversion factors from ADC counts to energy. To compare data and simulation, all 
inoperable or noisy channels in the SVT and ECal were disabled in both data and simulation so that 
any efficiency or bias that affect the data should be reflected in the simulation. 
Iteratively, conversion coefficients for each crystal were adjusted until the $E$/$p$ ratio in data and 
simulation were similar. The distribution of the $E$/$p$ ratio in data and simulation are compared in 
Figure~\ref{fig:gains}. 
The peak of the distribution, at $E/p\sim0.7$, gives the sampling fraction of the ECal, the fraction of the 
incident particle energy measured in the cluster. The width of the distribution indicates the energy 
resolution, which is worse than the required $5\%/\sqrt{E}$ for HPS due to high thresholds. 
The width in data is greater than that in simulation due to non-uniformity of the calibration of different parts 
of the ECal. 

The \Aprime{} trigger in HPS is relatively insensitive to the energy of the clusters and this level of 
performance would be adequate. However, improvements are needed to achieve the expected energy 
resolution in HPS. The noise and and thresholds need to be closer to 10~MeV and a 
more elaborate calibration technique needs to be employed to suppress the large tails in the $E$/$p$ 
distribution further. In addition, the fraction of working channels needs significant improvement.

\subsection{Trigger performance}
\begin{figure}[ht]
\begin{center}
	\includegraphics[width=7cm]{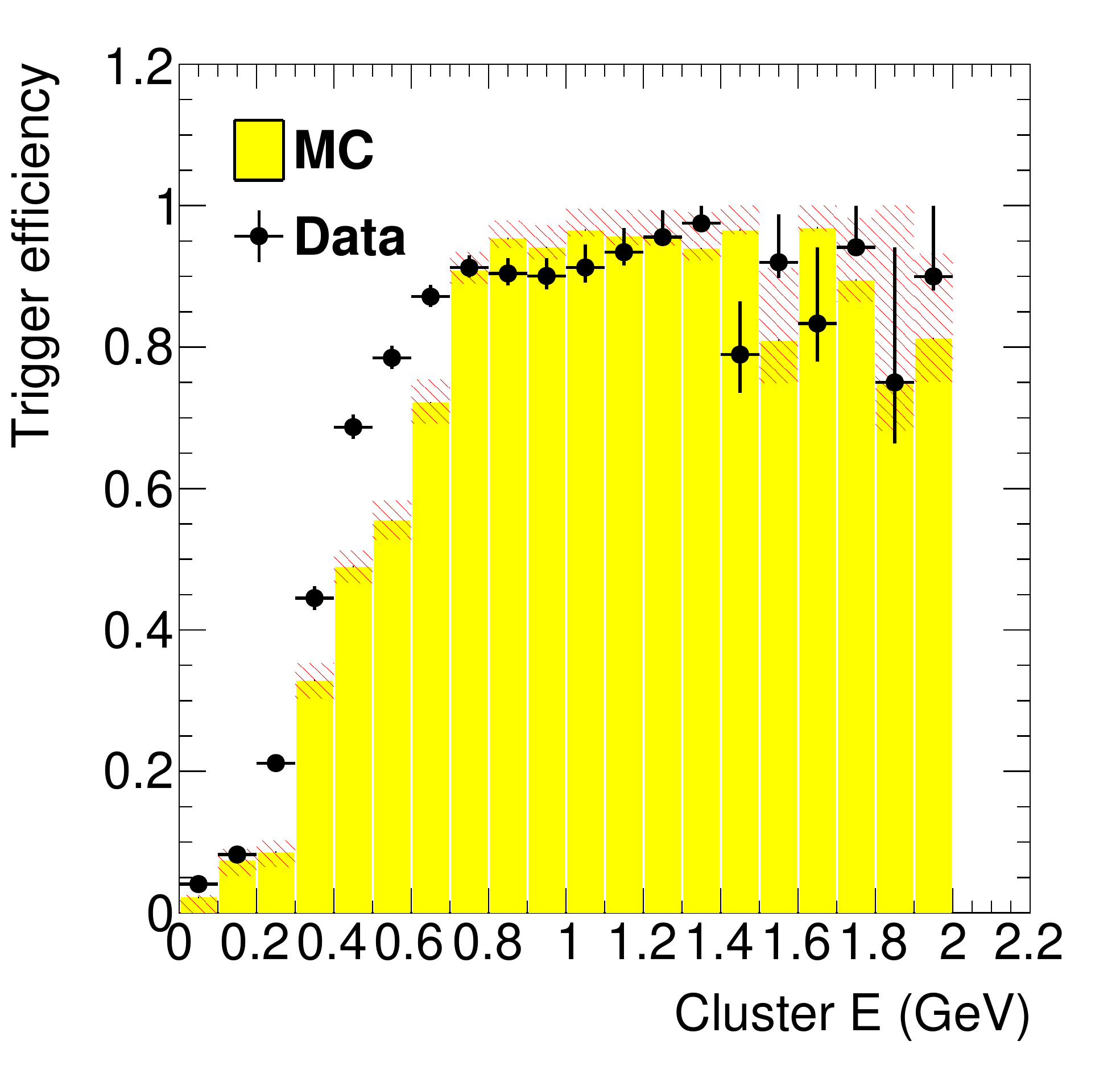}
	\caption{\small Trigger efficiency in both halves of the ECal for data and simulation as a 
	function of cluster energy.
\label{fig:turnon}}
\end{center}
\end{figure}

As described above in Sec.~\ref{sec:triggerdaq}, the energy from each crystal is determined slightly 
differently in the trigger and in the readout. The trigger performance was studied by simulating the 
trigger for each event and comparing to how the events were actually triggered.
To eliminate trigger bias, we use a tag and probe method: to study the trigger performance in one half 
of the ECal, we select 
events which triggered the other half and where there was exactly one probe cluster in the ECal half 
under study. We then measure trigger efficiency as the fraction of tagged events that fired the trigger in 
the probe half as a function of the probe cluster energy, shown in Figure~\ref{fig:turnon}. 
The trigger turn-on is slow and reaches an uneven plateau at about 700~MeV for two reasons;  
gain variations between different crystals lead to the threshold variations and the nonlinearity of 
the time-over-threshold integral means that the effective threshold is higher for clusters that span 
multiple crystals. The effective trigger threshold is therefore dependent on position and energy of 
the particle as well as cluster multiplicity. For HPS the trigger cluster threshold will be set at a lower 
value. 

As a cross-check we simulate the FADC trigger path by converting from readout hits (with fixed-size 
window integration) to trigger hits (time-over-threshold integration). The CTP clustering 
algorithm and the trigger decision from the SSP are simulated before we compare the trigger decision 
and trigger time to what was reported by the actual trigger. For every event, the trigger reports the 
trigger decision as a bit mask (top half, bottom half or both) and the time the trigger fired.
The turn-on from the trigger threshold was measured to be 1280 in units of ADC counts as expected. 
The threshold was not perfectly sharp because of uncertainties in the conversion from readout to trigger 
hits described above, but based on comparisons with simulation we found that the 
trigger worked exactly as specified.

\subsection{Trigger rate comparisons}
Trigger rates observed in the HPS Test Run are dominated by \ee{} pairs produced in the 
converter. In simulated events, the rate of triggers depend on the modeling of the pairs' angular 
distribution and the subsequent multiple Coulomb scattering in the converter. Rates from different 
converter 
thicknesses are used to study the varying multiple Coulomb scattering contribution (pair production 
angular distribution is constant).  These rates are compared with those predicted by the standard Geant4~\cite{Agostinelli2003250} approximation for multiple 
scattering and the more accurate model used by EGS5~\cite{egs5}. 
\begin{table}
{\small
\begin{tabular}{|l|c|c|c|}
\hline
\bf Converter (\%~$X_0$) & \bf 1.60 & \bf 0.45 &	\bf 0.18 \\
\hline
EGS5 &	1162 $\pm$ 112 &	255 $\pm$ 28 &	94 $\pm$ 17	\\
\hline
Geant4 & 2633 $\pm$ 250 & 	371 $\pm$ 38 &	114 $\pm$ 18 \\
\hline
Observed 	& 1064 $\pm$ 2 & 196 $\pm$ 1 &	92 $\pm$ 1 \\						
\hline
\end{tabular}
\caption{ Observed and predicted event rate (in Hz) normalized to 90~nA for three different converter 
thicknesses. The uncertainty on the prediction includes systematic uncertainties from ECal alignment, background normalization, beam current normalization and limited statistics in the simulation.
\label{results}}
}
\end{table}
Restricting to a well calibrated region of the ECal and to clusters with energy above the trigger turn-on, 
we see agreement with the rates predicted by the \egs{} simulation program after subtracting the 
``no converter'' background , see Table~\ref{results}.
This gives further confidence that the dominant source of background occupancy for HPS, multiple 
Coulomb scattered beam electrons, is well 
described.


\section{Summary and outlook}
The HPS Test Run experiment, using a simplified version of the apparatus planned for the full HPS 
experiment in a parasitic photon beam, demonstrated the feasibility of the detector technologies 
proposed for the silicon vertex tracker, electromagnetic calorimeter, and data acquisition systems. 
Performance from each of these subsystems has been shown to be adequate to conduct the full 
experiment successfully with some identified improvements. Studies of multiple Coulomb scattering tails of electrons and positrons from 
photon conversions further backs expectations from simulation, giving credence to estimates of the 
detector backgrounds expected in electron beam running for HPS.

\section{Acknowledgements}
The authors are grateful for the support from Hall~B at JLab and especially the Hall~B engineering 
group for support during installation and decommissioning. They also would like to commend the 
CEBAF personnel for good beam performance, especially the last few hours of operating CEBAF6. 
The tremendous support from home institutions and supporting staff also needs praise from the 
authors. 

Work supported by the U.S. Department of Energy under contract number DE-AC02-76SF00515, 
the National Science Foundation,  
French Centre National de la Recherche Scientifique and 
Italian Istituto Nazionale di Fisica Nucleare. 
Rouven Essig is supported in part by the Department of Energy Early Career research program 
DESC0008061and by a Sloan Foundation Research Fellowship. Authored by Jefferson Science 
Associates, LLC under under U.S. Department of Energy contract No. DE-AC05-06OR23177.





\bibliographystyle{model1-num-names}
\bibliography{hps-testrun-nim}







\end{document}